\documentstyle[dina4p,12pt,epsfig,rotating]{article}

\newcommand{\gp}{$\gamma p$ }
\newcommand{\degree}{\rm \circ}

\newcommand{\psiprime}{$\psi'$}
\def\gsim{\,\lower.25ex\hbox{$\scriptstyle\sim$}\kern-1.30ex%
\raise 0.55ex\hbox{$\scriptstyle >$}\,}
\def\lsim{\,\lower.25ex\hbox{$\scriptstyle\sim$}\kern-1.30ex%
\raise 0.55ex\hbox{$\scriptstyle <$}\,}

\newcommand{\ptrt}{$p_t^2$}

\newcommand{\vdmprp}{$(m^2_{\psi}/(Q^2+m^2_{\psi}))^2$}
\newcommand{\jmm}{$J/\psi\ra\mu^+\mu^-$}
\newcommand{\jee}{$J/\psi\ra e^+e^-$}

\newcommand{\ra}{\rightarrow }

\newcommand{\cs}{cross section}

\newcommand{\pdiss}{proton dissociation}
\newcommand{\cm}{center of mass}

\newcommand{\colsing}{colour singlet}
\newcommand{\coloct}{colour octet}
\newcommand{\colsingmodel}{\colsing\ model}

\newcommand{\picb}{\mbox{pb}^{-1}}

\newcommand{\ceme}{\,\mbox{cm}}
\newcommand{\nb}{\,\mbox{nb}}

\newcommand{\MeV}{\,\mbox{MeV}}

\newcommand{\GeV}{\,\mbox{GeV}}
\newcommand{\GeVt}{\,\mbox{\GeV$^2$}}
\newcommand{\gmt}{\,\mbox{\GeV$^{-2}$}}

\def\3{\ss}
\begin{document}
\sloppy
\newcommand{\wgp}{$W_{\gamma p}$ }
\newcommand{\mx}{$M_X$} 
\newcommand{\epcs}{$ep$ cross section}
\newcommand{\pgf}{photon gluon fusion}
\newcommand{\jpsi}{$J/\psi$}
\newcommand{\jpsiw}{J/\psi} 
\newcommand{\wgpw}{W_{\gamma p}} 
\newcommand{\ep}{positron proton}
\newcommand{\ee}{$e^+e^-$} 
\newcommand{\mm}{$\mu^+\mu^-$}

\begin{titlepage}
{\tt DESY 96-037} \\
{\tt March 1996}\\
\vspace{5cm}
\begin{center}
\begin{Large}
\boldmath
\bf{Elastic and Inelastic Photoproduction \\
of \protect\jpsi\ Mesons at HERA\\}
\unboldmath 

\vspace*{2.cm}
H1 Collaboration \\

\end{Large}

\vspace*{1cm}

\end{center}

\vspace*{1cm}

\begin{abstract}
\noindent
Results on \jpsi\ production in $e p$ interactions
in the H1 experiment at HERA are presented.
The \jpsi\ mesons are produced by almost real photons ($Q^2\approx 0$)
and detected via their leptonic decays. The data have been taken in 1994 and
correspond to an integrated luminosity of $2.7\,\mbox{pb}^{-1}$.
The $\gamma p$ cross section for elastic
\jpsi\ production is observed to increase strongly with the
\cm\ energy.
The cross section for diffractive $J/\psi$ production with proton dissociation
is found to be of similar magnitude as the elastic cross section. 
Distributions of transverse momentum and decay angle
are studied and found to be in accord with a diffractive production
mechanism.
For inelastic \jpsi\ production the total $\gamma p$ cross section, 
the distribution of transverse momenta, and the elasticity of the \jpsi\ 
are compared to NLO QCD calculations in a colour singlet model 
and agreement is found. 
Diffractive \psiprime\ production has been observed 
and a first estimate of the ratio to 
\jpsi\ production in       
the HERA energy regime is given.

\end{abstract}
\end{titlepage}

\begin{sloppy}
%--Status: 18/12/95
 S.~Aid$^{14}$,                   %HAM2-PD      8/93        Aid
 V.~Andreev$^{26}$,               %LPI -PD                  Andreev
 B.~Andrieu$^{29}$,               %ECPL-PD                  Andrieu
 R.-D.~Appuhn$^{12}$,             %DESY-LEFT  10/95         Appuhn
 M.~Arpagaus$^{37}$,              %ZUTH-LEFT   4/95         Arpagaus
 A.~Babaev$^{25}$,                %ITEP-PD                  Babaev
 J.~B\"ahr$^{36}$,                %ZEUT-PD                  Baehr
 J.~B\'an$^{18}$,                 %KOSI-PD                  Banj
 Y.~Ban$^{28}$,                   %ORSa-ST                  Bany
 P.~Baranov$^{26}$
% ...
\end{sloppy}    
%
%
%
%     H1 Institutes as appearing on publications
% ...
$ ^{12}$ DESY, Hamburg, Germany$^a$ \\
% ...
$ ^{14}$ II. Institut f\"ur Experimentalphysik, Universit\"at Hamburg,
                                     Hamburg, Germany$^ a$  \\
% ...
$ ^{18}$ Institute of Experimental Physics, Slovak Academy of
                Sciences, Ko\v{s}ice, Slovak Republic$^ f$\\
% ...
$ ^{25}$ Institute for Theoretical and Experimental Physics,
                                                 Moscow, Russia \\
 $ ^{26}$ Lebedev Physical Institute, Moscow, Russia$^ f$ \\
% ...
$ ^{28}$ LAL, Universit\'{e} de Paris-Sud, IN2P3-CNRS,
                            Orsay, France\\
 $ ^{29}$ LPNHE, Ecole Polytechnique, IN2P3-CNRS,
                             Palaiseau, France \\
% ...
$ ^{36}$ DESY, Institut f\"ur Hochenergiephysik,
                              Zeuthen, Germany$^ a$\\
 $ ^{37}$ Institut f\"ur Teilchenphysik,
          ETH, Z\"urich, Switzerland$^ i$\\
% ...
$ ^a$ Supported by the Bundesministerium f\"ur
        Forschung und Technologie, FRG,
        under contract numbers 6AC17P, 6AC47P, 6DO57I, 6HH17P, 6HH27I,
        6HD17I, 6HD27I, 6KI17P, 6MP17I, and 6WT87P \\
% ...
$ ^f$ Supported by the Deutsche Forschungsgemeinschaft\\
% ...
$ ^i$ Supported by the Swiss National Science Foundation\\
\vfill
\clearpage

\section{Introduction}

The high center of mass (cms) energy of $\sqrt{s}\sim 300\GeV$ 
available at the electron proton collider
 HERA has led to a renewed interest in the study of $J/\psi$
production in lepton proton scattering.
The production \cs\ is dominated by photoproduction, i.e.
by the interaction of almost real photons,
$Q^2 \approx 0$, where $Q^2$ is the negative four momentum transfer
squared to the scattered lepton.
The topic of this paper is an analysis of photoproduction of \jpsi\ mesons 
in {\em elastic} and {\em inelastic} processes.

Several mechanisms have been suggested to describe photoproduction
of \jpsi\ mesons
$$ \gamma p\ra J/\psi\,X.$$
For the description of 
the {\em elastic} (or {\em exclusive}) process where $X=p$, a 
diffractive mechanism has been proposed by Donnachie and Landshoff~\cite{DL1}.
The mechanism is based on Regge phenomenology where \jpsi\ production 
is mediated by pomeron exchange (Fig. 1(a)) and can be extended to include 
also {\em proton dissociation} (Fig. 1(b)).
Measurements at HERA are expected to shed new
light on the diffractive production mechanism and the nature 
of the pomeron~\cite{DL2}.\\
At lower \cm\ energies pomeron exchange was successfully  used in the 
framework of the vector dominance model (VDM)~\cite{VMD,Bauer}
to describe the production of {\em light} vector mesons.
The predicted slow variation with energy of the elastic \cs\ 
was confirmed and an exponentially 
falling $t$ distribution was found, where $t$ is the 
squared four momentum transfer to the scattered proton. 
When applied to $J/\psi$ meson production at low \cm\ energies
the prediction of the 
VDM model was found to be more than an order of magnitude 
above the data
but other
features like energy dependence and $t$ distributions agreed within errors.

Attempts have been made to describe {\em elastic} \jpsi\ production
in perturbative QCD. In the approach 
by Ryskin~\cite{Ryskin} the interaction between the proton and the charm 
quark is mediated by the
exchange of a gluon ladder (Fig. 1(c)). Non perturbative effects are included 
in the gluon distribution of the proton which enters the production cross 
section quadratically. 
A measurement of elastic $J/\psi$ production  could therefore be 
a sensitive probe of the gluon density in the proton. 
Recently  higher order effects have been calculated 
in this model and a comparison to preliminary HERA data~\cite{Bxl_papers} 
was carried out~\cite{Ryskin_et_al}. The Ryskin model predicts 
a different energy dependence of the elastic $J/\psi$ cross section
than the Donnachie-Landshoff approach. 
The energy dependence in the Ryskin model is coupled to the low $x$ behaviour 
of the gluon density in the proton.
Using a gluon density increasing towards low $x$
which describes recent measurements of the structure 
function $F_2$ at HERA\cite{h1f2}, 
results in a fast increase of the \cs\ for elastic photoproduction of 
\jpsi~\cite{Ryskin_et_al}.

In {\em elastic}
 \jpsi\ production  at small momentum transfer the \jpsi\ meson 
 retains approximately the full photon energy 
($z \approx 1$ with $z=E_{\psi}/E_{\gamma}$ in the proton rest system).
\jpsi\ production {\em with proton dissociation}
although strictly speaking an inelastic process
also leads to $z$ values close to 1.
In contrast, {\em inelastic} processes will have 
$z$ values below 1 and a high mass 
hadronic state is formed. 
The principal inelastic process proposed is the photon gluon fusion mechanism 
where the photon emitted by the incoming lepton interacts with a gluon from 
the proton via the charm quark (Fig. 1d). 
This mechanism is calculable in perturbative QCD due to the hard 
scale given by the mass of the charm quark. 
Attempts have been made to determine the gluon 
density 
in the proton from this reaction
in several fixed target 
experiments~\cite{aemc83,cemc92,ccnmc91}.
In the colour singlet model~\cite{Berger_Jones} for photoproduction of \jpsi,
 which is based on this 
picture, the formation of a $J/\psi$ state is
accompanied by the emission of a hard gluon. Comparing predictions of
the colour singlet model to data a discrepancy in absolute magnitude 
was found which was attributed to missing higher order calculations 
(``K--factor"). Subsequently several improvements have been proposed which 
led to better agreement with the data~\cite{jung_kr}. 
Recently, complete next-to-leading order (NLO)
calculations have been performed~\cite{kraem1,kraem2}
and compared successfully  to fixed target data and to 
preliminary HERA data.

\begin{figure}[tb]
\vspace*{13pt}
\setlength{\unitlength}{1cm}
\begin{picture}(14.0,8.0)
\put(-1.0,-11.0){\epsfig{file=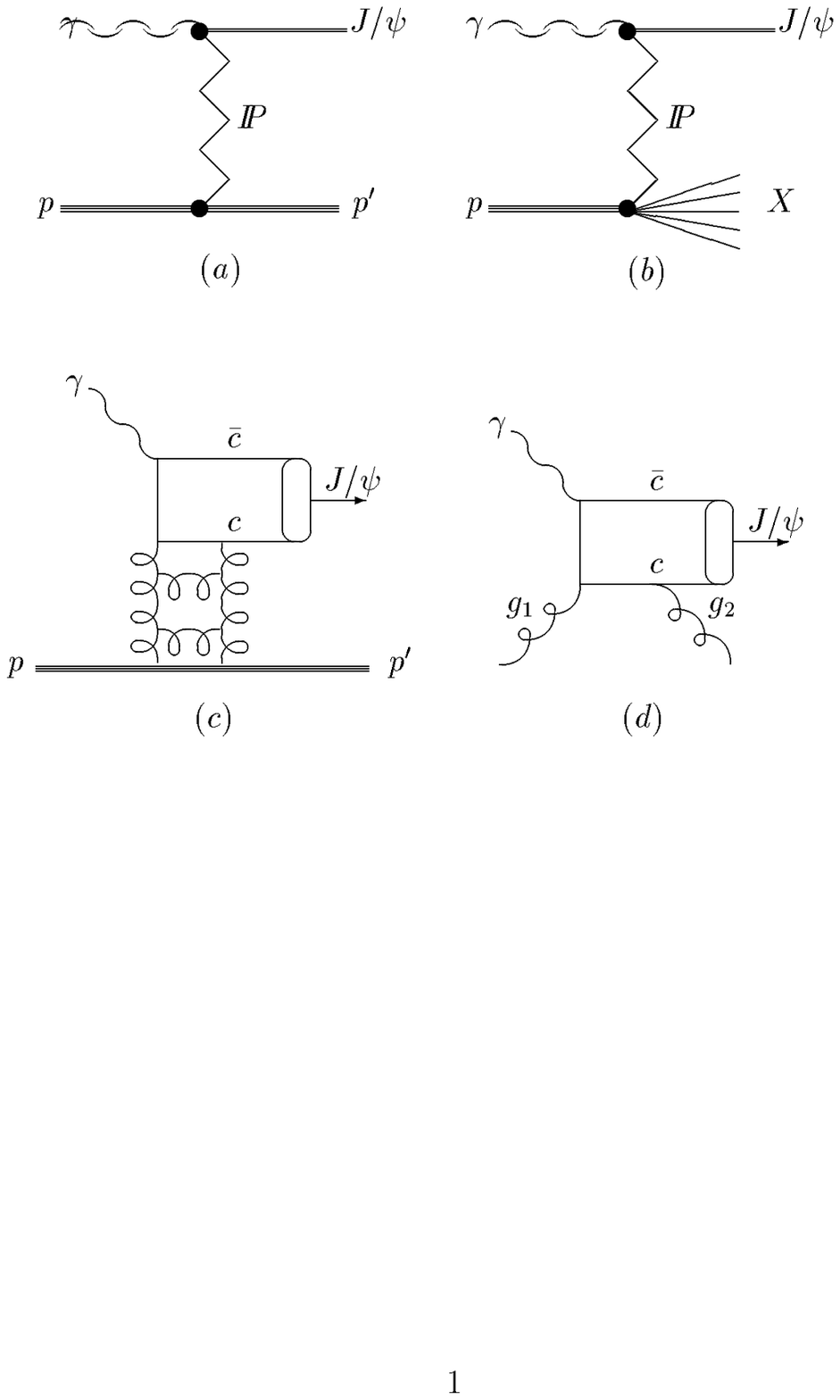}}
\end{picture}
\vspace*{13pt}
\caption{\sl\jpsi\ production mechanisms: $(a)$ elastic \jpsi\
production via pomeron exchange; $(b)$~
diffractive proton dissociation;
$(c)$ elastic \jpsi\ production in perturbative QCD: exchange of a gluon
ladder~\protect\cite{Ryskin};
$(d)$ photon gluon fusion model for inelastic \jpsi\ production
(\protect\colsingmodel~\protect\cite{Berger_Jones}).}
\end{figure}

Motivated by TEVATRON data on inelastic \jpsi\ and \psiprime\ 
production~\cite{teva} which may require colour octet 
contributions to explain  the measured cross sections, calculations have been 
performed for possible colour octet contributions in 
photoproduction at HERA~\cite{cacci,amund}. Both groups
attempt to estimate the \coloct\ contributions 
to \jpsi\ mesons at high $z$; in~\cite{cacci} the contribution to 
{\em inelastic} \jpsi\ production is estimated.
       
We present here an analysis of elastic and inelastic $J/\psi$ production in 
positron proton collisions near
$Q^2 = 0$  for $\gamma p$ cms
energies up to $150\GeV$. 
The analysis is based 
on $J/\psi$ decays to leptons, $J/\psi \rightarrow
\mu^+\mu^-$ or $J/\psi \rightarrow e^+e^-$. 
The data were collected with the H1 detector at HERA and correspond to an 
integrated luminosity of
$\sim 2.7 \, \mbox{pb}^{-1}$. 

The {\em inelastic} process is analysed for the
first time in H1; preliminary results have also been shown by the ZEUS 
collaboration~\cite{Bxl_papers}. 
The analysis of the {\em elastic} process is an update of a
previous letter~\cite{jpsi1} where we have presented a measurement of 
$\sigma(\gamma p \rightarrow J/\psi+X)$ which showed
a strong increase of the cross section with $W_{\gamma p}$, the photon
proton \cm\ energy, compared to  experiments at
lower cms energy. This increase was  faster than expected
from the  Donnachie-Landshoff prediction.
At that time contributions from processes with
proton dissociation could however not be excluded completely. 
With the increased statistics
now available and an improved analysis method these inelastic 
processes can be efficiently recognised. 
The fast increase of the elastic \cs\ with \wgp\ was also observed by the ZEUS 
collaboration~\cite{zeus1,Bxl_papers}.

The paper is organised as follows.
After a brief introduction of the kinematics (section 2),   
the experimental conditions
and the pre-selection of the data sample are described in section 3.
The analysis of the {\em elastic} \jpsi\ samples follows in section 4,
which includes 
the energy dependence of the total \gp\ \cs\ and the distribution of \ptrt, 
the transverse momentum of the \jpsi.
The second part of section 4 contains the analysis of the {\em proton dissociation} 
process and the decay angle distribution of diffractively produced 
\jpsi\ mesons.
In section 5 the selection of the inelastic events is discussed and 
results are presented including the energy dependence of the \gp\ \cs, 
the distributions of \ptrt\  and of the elasticity $z$ of the \jpsi.  
\newline 
A first cross section for \psiprime\ production in the HERA energy range
is given in the last section.

\section{Kinematics} 
\label{formulas}

The variables used for the 
description of $J/\psi$ photoproduction at HERA are:

\begin{eqnarray}
Q^2=-q^2&=&-(k-k')^2\\
       t&=&(P_p-P')^2\\
       s&=&(P_p+k)^2\\
       y&=&\frac{P_p \cdot q}{P_p \cdot k}\\ % \approx \frac{E_\gamma}{E_e}\\
W_{\gamma p}&=&\sqrt{(P_p+q)^2}
\end{eqnarray}

with the 4-momenta $P_p$ and $k$ of the incoming proton and
positron, $k'$ of the scattered positron, 
$q=k-k'$ of the exchanged photon, and  
$P'$ of the system X, which is identical to the proton for elastic
\jpsi\ production.

The Bjorken variable $y$ can for photoproduction be approximated by
$y\approx {E_\gamma}/{E_e}$, where $E_{\gamma}$ and $E_e$ are the energies 
of the exchanged photon and the incoming positron.
The variable $y$ can be computed from the observed 
final state using the method
by Jacquet--Blondel~\cite{blondel}:

\begin{equation} 
y = \frac{\Sigma\ (E -  P_z)}{2E_e} = \frac{(E - P_z)_{J/\psi} +
\Sigma_{\mbox{rest}}(E - P_z)}{2E_e}.
\label{yjb}
\end{equation}

The sum is over all visible particles, i.e. tracks and energy 
deposits in the calorimeter. In order to avoid double counting 
the calorimeter energy in a cylinder of radius $30\ceme$ 
around the extrapolated charged tracks is excluded. 
For events where only the two \jpsi\ decay leptons are  observed 
in the tracking detector the calorimeter
information is not used for the calculation of $y$. Analysing photoproduction
it is customary to use \wgp:
$$  \wgpw^2=y\,s-Q^2+m_p^2 \approx y\,s$$
The separation of the events into an elastic and an inelastic sample
utilizes the topology of the event and
the variable $z$ which is defined as:

\begin{equation} 
z = \frac{P_p \cdot P_\psi}{P_p \cdot q}
\end{equation} 

where $P_{\psi}$ is the four vector of the $J/\psi$.
Using eqn. (4) $z$ can be expressed as:

\begin{equation}
z=\frac{y_{\psi}}{y} \qquad \mbox{with} \qquad
                      y_{\psi}= \frac{(E-P_{z})_{J/\psi}}{2\,E_{e}},
\label{yjbz}
\end{equation}

For elastic events $y_{\psi}=y$ and thus $z=1$.

The \epcs\ and the  \gp\ \cs\ are  related by:

\begin{equation} 
  \sigma (ep\ra e\,J/\psi\,X) =
  \int_{y_{min}}^{y_{max}} dy \int_{Q^2_{min}(y)}^{Q^2_{max}} dQ^2
        \cdot
  f_{\gamma/e}(y,Q^2) \cdot
  \sigma_{\gamma^* p}(Q^2,y)
\label{sepgp}
\end{equation}

where the \cs\ on the left side is the measured $ep$ \cs\ covering 
a range $Q^2_{min}\leq Q^2\leq Q^2_{max}$ and determined in bins of $y$.
The kinematical minimum for a given value of $y$ is $Q^2_{min}$,
 and $Q^2_{max}$ is
the effective upper limit of the selected event samples: 

$$ Q^2_{min}=m_e^2\,\frac{y^2}{1-y};\qquad \;\;\;\qquad
                                 Q^2_{max}=4\GeV^2.$$

The flux of {\em transverse} photons is~\cite{rfflux}:

$$ f_{\gamma/e}(y,Q^2) = \frac{\alpha}{2 \pi} \frac{1}{y\,Q^2} \cdot 
  \left ( 1 + (1-y)^2 - \frac{2 m^2_e y^2}{Q^2} \right ).$$

The longitudinal photon flux amounts to $2\%$ of the transverse flux
for the given kinematical range.
Since the photon flux decreases 
rapidly with $Q^2$ and $y$, the weak $Q^2$ and $y$ dependences of 
$\sigma_{\gamma^* p}(Q^2,y)$ play only a minor role and 
the photoproduction \cs\  $\sigma (\gamma p\ra J/\psi\,X)$
can in first approximation be identified
with $\sigma_{\gamma^* p}(Q^2,y)$. It is then obtained as: 

$$  \sigma (\gamma p\ra J/\psi\,X)=
                  \sigma (ep\ra e\,J/\psi\,X)/\Phi_{\gamma/e} $$

where $\Phi_{\gamma/e}$ is the photon flux integrated over  $Q^2$ and $y$.
The correction for this approximation is small; it can be 
taken into account as a small change of the \wgp\ value at which the 
\cs\ measurement is performed.

%%%%%%%%%%%%%%%%%%%%%%%%%%%%%%%%%%%%%%%%%%%%%%%%%%%%%%%%%%%%%%%%%%%%%%%
%%%%%EXPERIMENTAL           %%%%%%%%%%%%%%%%%%%%%%%%%%%%%%%%%%%%%%%%%%%
%%%%%%%%%%%%%%%%%%%%%%%%%%%%%%%%%%%%%%%%%%%%%%%%%%%%%%%%%%%%%%%%%%%%%%%

\section{Experimental Conditions}

The data were taken in 1994 with the H1 detector operating at the electron 
proton storage ring HERA, where positrons of 27.5\GeV\ collide 
with protons of 820\GeV.
In 1994 HERA was operated with 153 colliding positron and proton 
bunches.
The integrated luminosity 
used for this analysis is 
 $2.7\,\picb$ for the decay \jmm\  and 
 $2.0\,\picb$ for \jee.
The  H1 detector is described in~\cite{H1}. We repeat here the essential 
features of the detector components used for the analysis.

\subsection{The H1 Detector}       
\label{sdet}
The central tracking system is mounted concentrically around the
beamline and covers polar angles\footnote{H1 uses a right-handed
             coordinate system defined as follows:
the origin is at the nominal interaction point with the $z$--axis
pointing in the proton beam direction, hence the polar angle $\theta$ is
measured with respect to the proton beam direction.
The region of small polar angles is called ``forward".
%The $y$--axis points upwards. 
The plane perpendicular to the $z$ axis is
named $r-\phi$ plane.}  between $20^{\degree}$ and
$160^{\degree}$. Measurements of charge and momenta of charged particles are
provided by two coaxial cylindrical drift chambers 
(central jet chambers, CJC)~\cite{CJC}.
Two sets of cylindrical drift chambers for measurement of the
$z$-coordinate
and multiwire proportional chambers~(MWPC) 
for triggering are placed at two
radial positions. One set surrounds the beamline within the inner CJC
and
the other is mounted in between the two jet chambers.
The central tracking system is complemented by a forward
tracking system which covers polar angles 
$7^{\degree} \lsim\theta\lsim 25^{\degree}$.
In the
present analysis the forward tracker  is only used to detect events 
with tracks other than the $J/\psi$ decay leptons.

The tracking system is surrounded by a highly segmented 
liquid argon (LAr) sampling 
calorimeter~\cite{calo} with an inner electromagnetic section consisting 
of lead absorber
plates with a total depth of 20 to 30 radiation lengths and an 
outer hadronic section with steel absorber plates.
Polar angles between $4^{\degree}$ 
and $153^{\degree}$ are covered by the calorimeter and
the total depth is 
4.5 to 8 interaction lengths depending on the polar angle.
The backward region, $155^{\degree} \lsim \theta \lsim 176^{\degree}$, 
is covered by a lead
scintillator calorimeter where the scattered positron is detected for
$Q^2 \gsim 4 \GeVt$. 
 
The magnetic field of 1.15~T is produced by a superconducting
solenoid surrounding the LAr calorimeter.

The iron flux return 
yoke surrounding the superconducting solenoid is instrumented
with limited streamer tubes to provide muon identification; it is
segmented into 10 iron plates of 7.5~\ceme\ thickness and instrumented 
with up to 16 layers of streamer tubes. Muon tracks are 
reconstructed in the region $4^{\degree}\lsim \theta \lsim 171^{\degree}$ 
with a spatial resolution of the order of 1\ceme.  

In addition to this central muon detector there is a toroidal 
muon spectrometer outside the main H1 magnet
covering small polar angles. In the present analysis its driftchambers 
which cover $3^{\degree} \lsim \theta \lsim 17^{\degree}$,
are used for the recognition of events with proton dissociation.
For the same purpose
a system of scintillators -- the proton tagger -- is placed 24~m downstream
the proton beam around the beampipe, covering an angular range of 
approximately  $0.06^{\degree}\lsim\theta \lsim 0.25^{\degree}$.

The luminosity is measured using the radiative process $ep\rightarrow
ep\gamma$ where the photon is detected
in a luminosity monitor~\cite{lumi,H1}.

\subsection{Trigger and Data Processing}

\label{trigger}

The background rate at HERA is high, mainly due to interactions of the beam 
protons with gas in the beam pipe or surrounding material and  due to 
photoproduction of light quarks. 
Therefore a restrictive 
trigger is necessary which 
cannot use the distinctive signature of 
the scattered positron since for photoproduction the positron stays in 
the beampipe\footnote{The scattered positron is with a few exceptions 
not detected in the low angle tagger, because there was 
practically no overlap of its sensitive region ($y\gsim 0.3$) with the bulk of 
the \jpsi\ data which are at $y\lsim 0.25$.}.
 The trigger essentially has to rely on
the decay leptons of the \jpsi. 

Compared to our previous report~\cite{jpsi1} the trigger for muons and 
for electrons from \jpsi\ decays
has been improved resulting in an approximate 
efficiency for electron pairs of 50\% 
 and for muon pairs of nearly 60\% with a tolerable background rate. 

The following triggers are utilized:
 
\begin{enumerate}
\item {\bf Track triggers from Multi Wire Proportional 
           Chambers~(MWPC)}~\cite{zvtxtrig}
      demanding the origin of the tracks in the $z$ direction
      to be near the nominal 
      vertex region.

\item A {\bf coplanar track trigger} constructed from the MWPCs
         which demands
         exactly two tracks of transverse momentum $p_t\gsim 500\MeV$, 
         roughly coplanar
         with the beams.

\item A {\bf driftchamber track trigger }~\cite{rphitrig}
      demanding   one track in the CJC with a transverse
      momentum $p_t>450\MeV$ which originates  from the 
      interaction point in the plane perpendicular to the beams
      within $\sim 2\ceme$. 
\item {\bf Muon triggers} 
      demanding a penetrating particle
      detected in the central muon systems.  
\item A {\bf calorimetric trigger} 
      for low energy electromagnetic 
      clusters with $E>800 \MeV$
      roughly aligned with a track candidate in the proportional chambers.
\end{enumerate}

These triggers are combined such that 
each event class 
is triggered by at least two different trigger combinations 
which can be compared with each other for  determination of the 
efficiency.

For exclusive $J/\psi$ meson production in both leptonic decay modes
a combination of elements 1, 2 and 3 gives a trigger based purely on tracking 
chambers. A trigger for muons -- which is used for elastic and inelastic 
processes -- is based on track triggers 1 and 3 and the
muon trigger 4.
For electrons the combination of track triggers 1 and 3 with the calorimetric 
trigger 5  is used imposing a requirement of low track
multiplicity in the MWPCs therefore only sensitive to elastic and \pdiss\
processes.
       
\subsection{Track Selection and Lepton Identification}

The event selection starts from tracks found in the central drift 
chambers CJC which have been
associated to the primary $e^+p$ interaction point 
(vertex) 
in a constrained fit which helps to
increase
momentum resolution and to reduce pattern recognition ambiguities.    

{\em Electrons} are identified in the electromagnetic section 
of the LAr Calorimeter by linking a reconstructed drift chamber
track to a calorimeter cluster with energy $E_{cluster}>0.8\GeV$
and demanding the measured energy
to be comparable with the momentum ($E_{cluster}/p_{track}>0.7$).
The efficiency for electron identification is $\approx 85\%$ for
$p>0.8\GeV$.

A particle  is identified as a {\em muon} if the track in
the drift chamber is
either linked to a track element reconstructed in the central muon
detector or if it is identified as a minimum ionizing particle
in the LAr Calorimeter.
For the link  with a track in the central muon
system a drift chamber track is extrapolated taking 
into account the deflection in the magnetic field, the energy loss in the 
material of the detector and multiple scattering. 
The $\chi^2$ probability comparing the parameters
of the two tracks is required to be above $1\%$.
A muon signature in the LAr Calorimeter is defined by an  
energy deposit below $3 \GeV$ around the extrapolated track, 
at least three active cells in the hadronic part of the calorimeter, 
and the particle has to penetrate at least 90 \% of the  
calorimeter.
With these requirements the thresholds are 0.8 \GeV\ for the
identification of muons in the
LAr calorimeter ($\approx 75\%$ efficiency above $\sim 1\GeV$)  
and 1.5 \GeV\ in the central muon system ($\approx 80\%$ efficiency
above $\sim 2\GeV$).

\subsection{Event Simulation and Efficiency Determination}
\label{ghghjk}
 
Monte Carlo (MC) techniques are used
in order to determine the geometrical acceptance, 
the trigger and selection efficiencies.
Events are generated using  models and the detector
response is simulated in detail.
The simulated events are subjected to the same reconstruction
and analysis chain as the data.

Two models are used to simulate \jpsi\ production. Inelastic \jpsi\ 
production is simulated by the generator EPJPSI~\cite{jungws} which is based 
on the \colsing\ model in leading order.
Diffractive \jpsi\ production -- elastic and proton dissociation --
is modelled by the Monte Carlo generator
DIFFVM~\cite{benno}.

Diffractive events are generated with an energy dependence 
proportional to $W_{\gamma p}^{0.9}$, an exponential $t$ distribution 
$\sim \exp{(-b|t|)}$ with a fixed slope  
of $b=4\gmt$ in the elastic mode and $b=2\gmt$ 
for proton dissociation.   
The proton dissociation mode has an additional parameter, the mass \mx\ of 
the dissociated 
proton. Events are  generated with a  $1/M^2_X$ distribution for $M_X^2$ 
above $4\GeVt$. Below $4\GeVt$ 
the distribution is closely modelled to diffractive proton 
dissociation  data obtained from deuterium measurements~\cite{goulian}. 
The system $X$ is fragmented at masses above 2\GeV\ by treating 
it as a system of quark and diquark and using the 
Lund string model~\cite{lund}. 
Below 2\GeV\ it is treated 
as a nucleon resonance and decays into nucleon and pions according to the known
branching ratios.
  
The detector response in the Monte Carlo simulation is checked in detail
by comparing to the data.
After applying small overall correction factors the dependence of the 
efficiencies for track reconstruction, particle identification and triggering
on the polar angle and the momentum of the 
$J/\psi$ decay particles is well described by the simulation.
This tuned Monte Carlo simulation is then used for correcting the data
and remaining differences between data and Monte Carlo are included in the
systematic errors.

%%%%%%%%%%%%%%%%%%%%%%%%%%%%%%%%%%%%%%%%%%%%%%%%%%%%%%%%%%%%%%%%%%%%%%%
%%%%%%%%%%%%%%%%%%%%%%%%%%%%%%%%%%%%%%%%%%%%%%%%%%%%%%%%%%%%%%%%%%%%%%%
%%%%%%%%%%%%%%%%%%%%%%%%%%%%%%%%%%%%%%%%%%%%%%%%%%%%%%%%%%%%%%%%%%%%%%%

\subsection{Pre--selection of the $\boldmath{J/\psi}$ Samples}
\label{presel}

Elastic and inelastic \jpsi\   
events in general look quite different in the detector, 
elastic events having only the decay leptons measured
while inelastic events are characterized by additional tracks.
Initially a common selection aims to identify lepton pairs in the \jpsi\ mass
region irrespective of any other activity in the detector.
Two tracks are selected with a common origin at the beam interaction
point in the $r-\phi$ plane and momenta above $0.8\GeV$ in the range of
polar angles $20^{\degree}\leq \theta \leq 160^{\degree}$.
Both tracks have to be identified as muons or electrons.
The $z$-coordinate of the event vertex  has to be within $40\ceme$ 
of the average beam collision point.

For low multiplicity events, one of the main backgrounds is due to cosmic
ray muons, which are efficiently rejected by demanding  the angle between
the lepton candidates to be less than $177^{\degree}$.

Photoproduction events are selected by requiring 
no scattered positron be visible in the calorimeter,
i.e. below a polar angle of $176^{\degree}$. 
Rejecting  energy clusters above $8\,\GeV$ restricts the
photon virtuality to $Q^2\lsim 4\,\GeVt$.

%%%%%%%%%%%%%%%%%%%%%%%%%%%%%%%%%%%%%%%%%%%%%%%%%%%%%%%%%%%%%%%%%%%
%
%   E L A S T I C
%
%%%%%%%%%%%%%%%%%%%%%%%%%%%%%%%%%%%%%%%%%%%%%%%%%%%%%%%%%%%%%%%%%%%

\section{Analysis of the Elastic and Proton Dissociation Processes}
\label{twotr}

Starting from the pre-selected data a {two track sample} is 
selected for the determination of the \cs\ for elastic \jpsi\ 
production and for proton dissociation processes.
The selected lepton pairs are required to be the {\em only tracks}
coming from the beam interaction point in the
sensitive region of the tracking detectors 
($7^{\degree} \lsim \theta \lsim 165^{\degree}$). 

\begin{figure}[tbp]
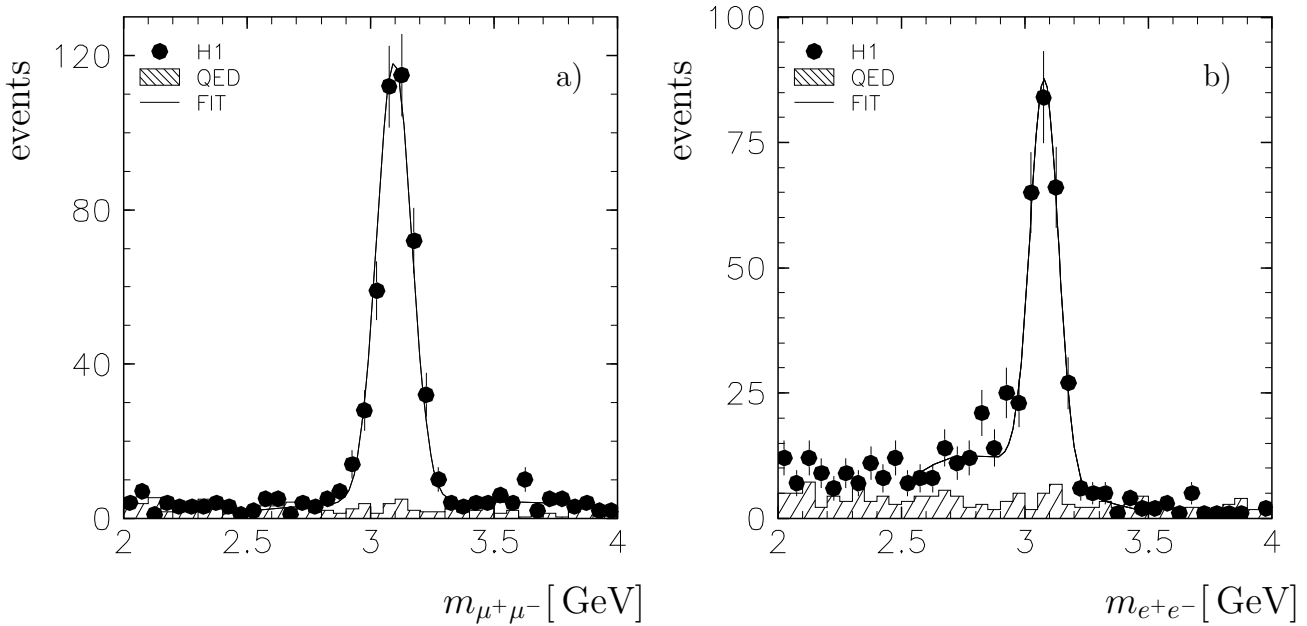

\vspace*{13pt}
\setlength{\unitlength}{1cm}
\begin{picture}(14.0,7.5)
\put(-1.6,-1.6){\epsfig{file=PAPER.ELAMASS.MU,height=33cm,angle=90}}
\put(0.2,5.5){\begin{rotate}{90}\large events\end{rotate}}
\put(5.7,-0.5){\large $m_{\mu^+\mu^-}[\GeV]$}
\put(7.2,6.5){a)}
\put(15.8,6.5){b)}
\put(7.1,-1.6){\epsfig{file=PAPER.ELAMASS.EL,height=33cm,angle=90}}
\put(9.0,5.5){\begin{rotate}{90}\large  events\end{rotate}}
\put(14.5,-0.5){\large $m_{e^+e^-}[\GeV]$}
\end{picture}
\vspace*{13pt}
\caption{\sl Mass distribution for $\mu^+\mu^-$ (a) and
        $e^+e^-$ pairs (b) for the two track 
        selection above $2 \GeV$. The curves
        are fits of a Gaussian plus a polynomial background to the
        $J/\psi$ mass region. The shaded histogram shows the 
        contribution of QED lepton pairs. For $\mu^+\mu^-$ the 
        maximum of the fit is at $(3.10 \pm 0.01)\GeV$ with a width
        of 76~MeV. For $e^+e^-$ 
        the maximum is at  $(3.08 \pm 0.02)\GeV$ and the width is  77~MeV.
        For both cases the detector simulation yields a width of 65~MeV.}
\label{ELAMASS}
\end{figure}

The reconstructed invariant mass of the lepton pairs for these events
is shown  in Fig. \ref{ELAMASS}a) and b)  for electrons and
muons, respectively. 
A cut around the nominal mass $\left|m_{\ell^+\ell^-} - m_{\psi}\right|
< 225 \MeV $ is applied.
This yields about 400 (350) $J/\psi$ candidates for muon (electron) pairs.

% MASS PLOTS ELASTIC
%%%%%%%%%%%%%%%%%%%%%

As can be seen from Fig. \ref{ELAMASS} the background below the 
\jpsi\ mass peak for this two track sample is low. 
It can be described by lepton pairs produced via a two photon process,
where a photon is emitted by the incoming positron and 
by the proton.
This process is  calculable in QED. The calculation
is implemented in the Monte
Carlo generator LPAIR~\cite{lpair}, the result of which is also shown 
in Fig. \ref{ELAMASS}.  

A subtraction is made for the background below the resonance, 
approximately 5\% for
\jmm\ and 12\% for \jee. The  number for 
electron pairs is larger because it 
contains two effects in addition to the QED lepton pairs.
There is an additional background of 
misidentified hadrons with a decreasing mass spectrum which has to be 
subtracted; on the other hand there is a small loss of signal events
due to the low mass tail of the \jpsi\ peak.
The latter is due to radiation in the 
material of the detector according to the detector simulation.

At this stage the two track event sample is composed of elastic events 
and those with proton dissociation where the fragments 
of the excited proton are not visible in the tracking detector.
The events are then classified into two samples, one {\em with} 
and one {\em without} activity in the forward detectors. 
The sample with forward activity will be used in section~\ref{spdis}
to determine the \cs\ for \jpsi\ production with proton dissociation. 
For elastic \jpsi\ production 
the sample without signals in the forward detectors is used.

The forward detectors close to the proton beam (see section~\ref{sdet}) 
are the proton tagger,
the drift chambers of the forward muon spectrometer,
and the low angle region of the liquid argon calorimeter 
($4^{\degree}\lsim\theta \lsim 10^{\degree}$). They
are sensitive to particles originating from secondary interactions 
of the excited nucleon fragments with material of collimators and the 
beampipe. Their efficiency has been studied with data~\cite{jansen}.

The three detector systems have different thresholds as regards
the mass $M_X$ of the dissociated nucleon system $X$. 
The proton tagger is sensitive down to 
the lightest nucleon excitation, the forward
muon detector is sensitive for $M_X \gsim 1.5\GeV$ and the low angle region
of the LAr calorimeter for $M_X \gsim 3 \GeV$.

\subsection{Elastic {\boldmath\jpsi} Production}
\label{sect elas}

Monte Carlo studies show that after the two track selection, 
$85\%$ of the events with proton dissociation
are recognised in the forward detectors.
The elastic event sample is selected by 
rejecting those events containing an energy
deposit above $0.75\GeV$ in the forward LAr calorimeter, or having 
hits in the proton tagger or having more than one hit
pair in the forward muon system. The 
event numbers are listed in table~\ref{elamu}.

\newcommand{\rb}[1]{\raisebox{1.5ex}[-1.5ex]{#1}}
\newcommand{\sgppsip}{$\sigma(\gamma p\ra\jpsiw p)$}
\newcommand{\seppsip}{$\sigma(e p\ra e\jpsiw p)$}
\newcommand{\sgppsix}{$\sigma(\gamma p\ra\jpsiw X)$}
\newcommand{\seppsix}{$\sigma(e p\ra e \jpsiw X)$}

\newcommand{\fpdis}{$1-f_{p.diss.}$}
%%%% E L A S T I C   M U   M U 

\begin{table}[tbp]
\begin{center}
\begin{tabular}{|l|c|c|c|c|}
%%%%%%%%%%%%%%%%%%%%%%%%%%%%%%%%%%%%%%%%%%%%%%%%%%%%%%%%%%%%%
\hline
$W_{\gamma p}[\GeV]$ & 30--60 &60--90 & 90--120 & 120--150 \\ 
\hline
%%%%%%%%%%%%%%%%%%%%%%%%%%%%%%%%%%%%%%%%%%%%%%%%%%%%%%%%%%%%%
\multicolumn{5}{|l|}{\bf \boldmath $J/\psi \rightarrow \mu^+\mu^-$ } \\
\hline 
\ events     & &  &  &   \\ 
(bg. corrected)  &\rb{52.0$\pm$7.5}  & \rb{64.0$\pm$8.6} & \rb{80.0$\pm$10.0}  
       & \rb{37.0$\pm$6.6} \\ 
\hline
$\epsilon_{acc}$ & 0.512$\pm$0.020 & 0.844$\pm$0.034 & 0.669$\pm$0.027 &
0.390$\pm$0.016  \\
\fpdis & 0.88$\pm$0.12  & 0.87$\pm$0.11 & 0.88$\pm$0.11 & 0.87
$\pm$0.11 \\ 
$\epsilon_{selection}$ &0.486$\pm$0.039 & 0.396$\pm$0.032 & 0.485$\pm$0.039
& 0.581$\pm$0.046  \\ 
$\epsilon_{trigger}$ &0.371$\pm$0.030 & 0.490$\pm$0.039& 0.724$\pm$
0.058& 0.828$\pm$0.066  \\
\hline 
\seppsip [nb] &2.9$\pm$0.4$\pm$0.5 & 2.0$\pm$0.3$\pm$0.4 &
1.7$\pm$0.2$\pm$0.3 & 1.0$\pm$0.2$\pm$0.2  \\ 
%%%%%%%%%%%%%%%%%%%%%%%%%%%%%%%%%%%%%%%%%%%%%%%%%%%%%%%%%%%%%
\hline
\hline
\multicolumn{5}{|l|}{\bf \boldmath $J/\psi \rightarrow e^+e^-$}\\
\hline 
  events   &  &  & & \\ 
 (bg. corrected) & \rb{48.4$\pm$7.5} & \rb{62.5$\pm$8.5} & \rb{43.1$\pm$7.1} 
  & \rb{10.6$\pm$3.5} \\ 
\hline 
$\epsilon_{selection}$ &0.550$\pm$0.038 & 0.515$\pm$0.036 & 0.447$\pm$
0.031 & 0.223$\pm$0.016 \\
$\epsilon_{trigger}$ &0.474$\pm$0.033 & 0.550$\pm$0.039 & 0.684$\pm$
0.048 & 0.791$\pm$0.055 \\
\hline 
\seppsip [nb] &2.5$\pm$0.4$\pm$0.4 & 1.8$\pm$0.2$\pm$0.3&
1.5$\pm$0.2$\pm$0.3& 1.1$\pm$0.3$\pm$0.2\\ 
%%%%%%%%%%%%%%%%%%%%%%%%%%%%%%%%%%%%%%%%%%%%%%%%%%%%%%%%%%%%%
\hline
\end{tabular}
\end{center}
\caption{\sl { \bf Elastic {\boldmath $J/\psi$} production}: 
Efficiencies and cross sections for elastic $J/\psi$ production
measured via the leptonic decays to 
$\mu^+\mu^-$  and $e^+e^-$   are given in intervals
of $W_{\gamma p}$  ($Q^2\protect\lsim 4\GeVt$) in the acceptance region. 
The number of events (two track selection requiring {\bf no} forward tag)
is corrected for background
below the mass peak. The remaining proton 
dissociation  background $f_{p.diss.}$ is calculated by Monte Carlo.
The acceptance $\epsilon_{acc}$ describes the probability for both leptons to
be in the $\theta$ range $20^{\degree} \leq \theta \leq 160^{\degree}$. 
The selection efficiency $\epsilon_{selection}$ contains the efficiencies for
 track 
reconstruction, lepton identification and selection cuts.
The errors of  
acceptance, proton dissociation background, 
selection and trigger efficiency give
the systematic errors.
The first error of the $ep$ cross
section is statistical, the second one  systematic.}
\label{elamu}
\label{elael}

%%%% COMBINED CROSS SECTION

\begin{center}
\begin{tabular}{|l|c|c|c|c|}
\hline 
\multicolumn{5}{|c|}{$J/\psi \rightarrow l^+l^-$}\\
\hline 
$ W_{\gamma p}[\GeV]$ & 30--60 &60--90 & 90--120 & 120--150\\ 
\hline 
$W_{0}[\GeV]$ & 42 &72 & 102 & 132 \\ 
\hline
$\Phi_{\gamma/e}$ &0.0736 & 0.0370 & 0.0229 & 0.0149 \\ 
\hline 
\hline 
\sgppsip$^{\mu\mu}$[nb] &39.6$\pm$5.7$\pm$7.2  & 53.4$\pm$7.2$\pm$9.8
      & 76.2$\pm$9.5$\pm$13.9 & 67.3$\pm$11.9$\pm$12.3 \\ 
\sgppsip$^{ee}$[nb] &34.4$\pm$5.3$\pm$6.0 & 48.2$\pm$6.6$\pm$8.4&
63.8$\pm$10.4$\pm$11.1  & 71.0$\pm$23.4$\pm$12.4 \\ 
\hline 
\sgppsip [nb] &36.8$\pm$3.9$\pm$6.6&50.6$\pm$4.8$\pm$9.1 
&70.6$\pm$7.0$\pm$12.7&68.0$\pm$10.6$\pm$12.2  \\ 
\hline 
\hline
\sgppsix [nb] &23.0$\pm$3.2$\pm$4.0&63.5$\pm$5.8$\pm$11.4 
&62.7$\pm$7.4$\pm$11.2&128.9$\pm$19.5$\pm$23.2  \\ 
\hline
\end{tabular}
\end{center}
\caption{\sl {\bf Elastic {\boldmath $J/\psi$} production}
and  $J/\psi$ production with {\bf proton dissociation}:
 $\gamma p$ cross sections for the elastic 
production of $J/\psi$ mesons are given 
for decays to muons and electrons separately and also
combined.  In the last line the combined cross section for
\jpsi\ production with proton dissociation is given.
$W_0$ is the corrected bin center.
$\Phi_{\gamma/e}$ is the photonflux integrated over $Q^2$ and \protect\wgp. 
The first error of the  cross
section is the statistical error, the second one systematic.}
\label{elacomb}
\end{table}

The remainder of the proton dissociation background is subtracted 
statistically using the Monte 
Carlo simulation program DIFFVM of \jpsi\ production with  
proton dissociation. The simulated events are
normalized to the number of events with forward detector signal.
All three forward detectors independently
lead to the same normalization within errors.
In  table~\ref{elamu} the correction factor which is applied to account for
remaining proton dissociation background is indicated in the line
marked '$1-f_{p.diss.}$'.

Additional background from \psiprime\ 
decays is removed taking into account the measured fraction of 
\psiprime/$\psi$ production
of $\approx 20\%$ (see section~\ref{psiprime}) and applying the selection
cuts to the simulated cascade decays into \jpsi$+anything$.
A correction of $4\pm 2\%$ is applied.

The remaining event sample has been examined by a visual scan in order to 
detect unrecognised background. A small number of events has energy deposits in
the calorimeter which might indicate a background. This number is  
less than the background expected due to \psiprime\ decays
and no further correction is applied.

\subsubsection*{Cross section as function of {\boldmath $W_{\gamma p}$}}

The elastic $ep$ cross section  is evaluated 
in bins of $W_{\gamma p}$ in the range of 30 -- 150 \GeV.
The $Q^2$ range is limited to $Q^2\lsim 4\GeV^2$ with an average of 
$0.13\GeVt$. 
The \cs\ was calculated according to:

$$
  \sigma_{ep}=\frac{N_{bg.corr.}(1-f_{p.diss.})\,(1-f_{\psi'})}
               {\epsilon_{sel}\,\epsilon_{trigger}\,\epsilon_{acc}\,\cal L} $$

where the factors are the number of events  corrected for 
background below the mass peak $N_{bg.corr.}$, 
the background fractions of proton dissociation  and \psiprime\ 
$f_{p.diss.}$ and  $f_{\psi'}$, respectively,
the efficiencies for data selection $\epsilon_{sel}$
(including track reconstruction, selection cuts, and lepton identification)
efficiency for triggering $\epsilon_{trigger}$, the geometrical acceptance
$\epsilon_{acc}$, and the integrated luminosity $\cal L$.
The results for the $ep$ \cs\  are given in table~\ref{elamu}
taking into account the branching ratio
for one leptonic decay channel of $(6.0\pm 0.25)\%$. The $ep$ \cs\ can be 
converted into a \gp\ \cs\ using the relations of
section~\ref{formulas}.
The integrated photon flux is shown in table~\ref{elacomb}.
The center $W_0$ of the \wgp\ bins of $30\,\GeV$ width is calculated 
taking into account
a  $y$ dependence of the $\gamma p$ \cs\ corresponding to 
$\sim W_{\gamma p}^{0.9}$,
a $Q^2$ dependence of $\sigma_{\gamma^* p}$   
as given by the VDM propagator \vdmprp, and the longitudinal \cs\ to be
$Q^2/m_{\psi}^2$ times the transverse \cs. 
The difference to the uncorrected bin center is estimated to be $-3\GeV$.

\begin{figure}[tbp]
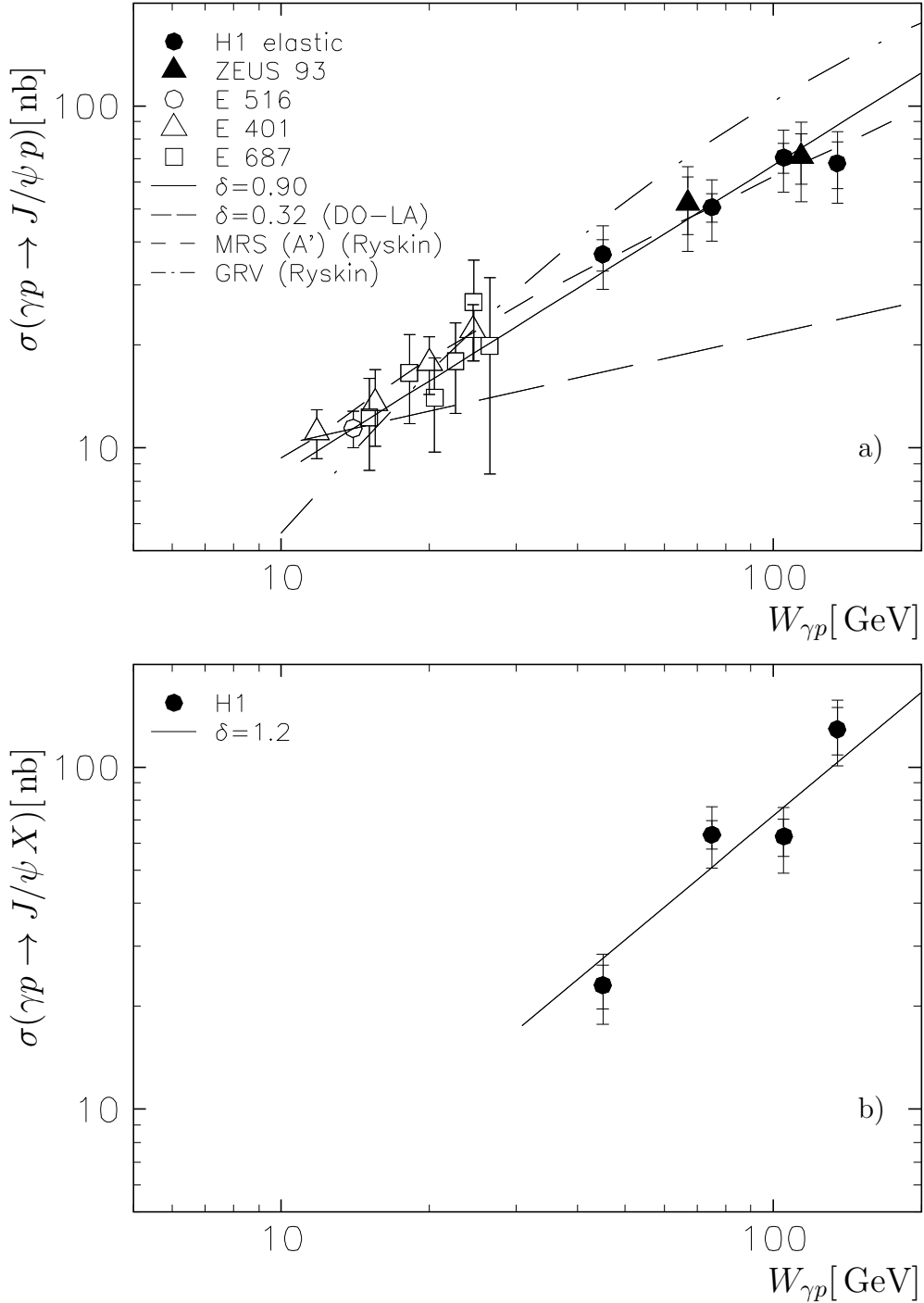

\vspace*{13pt}
\setlength{\unitlength}{1cm}
\begin{picture}(14.0,17.2)
\put(-0.75,7.5){\epsfig{file=PAPER.ELA.CROSS.WGP1,height=35cm,angle=90}}
\put(1.5,13.0){\begin{rotate}{90}\large $\sigma(\gamma p \rightarrow J/\psi \, p)
[\nb]$\end{rotate}}
\put(12.,9.0){\large $W_{\gamma p}[\GeV]$}
\put(-0.75,-2.0){\epsfig{file=PAPER.PD.WGP,height=35cm,angle=90}}
\put(1.5,3.0){\begin{rotate}{90}\large $\sigma(\gamma p \rightarrow J/\psi \, X)
[\nb]$\end{rotate}}
\put(12.,-0.5){\large $W_{\gamma p}[\GeV]$}
\put(13.3,11.5){a)}
\put(13.3,2.0){b)}
\end{picture}
\vspace*{13pt}
\caption{\sl {\bf a)} Total cross section for {\bf elastic} 
$J/\psi$ photoproduction. 
The inner
error bars of the HERA points are statistical, the outer ones
contain statistical and systematic errors added in quadrature. The data
at lower cms energies are from previous 
experiments \protect\cite{de687,ee40182,fftps}; they
were corrected with the new $J/\psi$ decay branching ratio
where necessary and
include systematic errors (added in quadrature).
A parametrisation of the energy dependence
as $W_{\gamma p}^{\delta}$ with $\delta=0.90$ (full curve)
and $\delta=0.32$ (long dashed) normalized to the E 516 point is shown.
Also shown is the result of 
calculations according to the Ryskin model~\protect\cite{Ryskin_et_al} using the MRSA'
(dashed) and GRV parametrisations (dash dotted)
of the gluon density. {\bf b)}
 Total cross section for 
$J/\psi$-production with {\bf proton dissociation}. 
The inner
error bars of the H1 points are statistical, the outer ones
contain statistical and systematic errors added in quadrature.
The full line represents a fit to $W_{\gamma p}^{\delta}$
with $\delta = 1.2 \pm 0.2$.}
\label{CROSS.WGP1}
\label{PD.WGP}

\end{figure}

Since the data based on \jpsi\ decays into
$\mu^+\mu^-$ and $e^+e^-$ 
agree within errors the $\gamma p$ cross sections have been averaged.
The resulting combined cross section is  shown  in Fig.~\ref{CROSS.WGP1} 
and listed in table~\ref{elacomb}. 
The data show a rise with
$W_{\gamma p}$ which in the HERA regime can be represented as
$\sigma_{\gamma p} \propto W_{\gamma p}^{\delta}$.
A fit to the H1 data yields $\delta = 0.64 \pm 0.13$,
where the error includes statistics and systematical effects.
In Fig.~\ref{CROSS.WGP1} also results by the ZEUS collaboration~\cite{zeus1} 
in the same energy range as the present data are shown, which agree
well.
Using also the ZEUS data and the measurements at 
lower energies~\cite{de687,ee40182,fftps} which are also shown in 
Fig.~\ref{CROSS.WGP1}, a combined fit
yields $\delta=0.90 \pm 0.06$.\\
The energy dependence expected from the Donnachie-Landshoff model~\cite{DL1} 
corresponds to $\delta=0.32$ (neglecting any shrinkage, see also next section)
and is also shown in Fig.~\ref{CROSS.WGP1}. It falls below the HERA data 
by more than a factor of 3 corresponding to more than 3
standard deviations if normalised at fixed target energies.

%ELASTIC CROSS SECTION
%%%%%%%%%%%%%%%%%%%%%%

The prediction of the QCD model due to Ryskin and including higher order
corrections~\cite{Ryskin_et_al}
is also compared with the data in Fig.~\ref{CROSS.WGP1}. 
The prediction depends quadratically on the gluon distribution 
taken at the scale $(Q^2+m_{\psi}^2)/4\approx 2.4\GeVt$, which can 
be parametrised as
$x\,g(x)\propto x^{-\lambda}$ at low values of $x$,
the fraction of the proton momentum 
carried by the gluon.
 For the present data 
$x\approx m_{\psi}^2/W_{\gamma p}^2\approx 10^{-3}$. 
Using the gluon distribution function from MRSA'~\cite{MRS}
corresponding to $\lambda \approx 0.2$,
gives good agreement with the data between 10 and 150\GeV. The
sensitivity to the gluon distribution is illustrated by comparison 
to the GRV fit~\cite{GRV} which corresponds to $\lambda \approx 0.3-0.4$.
The corresponding curve in Fig.~\ref{CROSS.WGP1} 
with the parameters used in~\cite{Ryskin_et_al}
yields a steeper energy dependence than the data.

\subsubsection*{Systematic errors}

\label{systerr}
A breakdown of the systematic error  of the \cs s
in tables~\ref{elamu} and \ref{elacomb} is given in table~\ref{tserr}.

The main contribution to the systematic error of elastic and \pdiss\ 
\cs s is due to the separation of these two event classes. The 
uncertainty is estimated to be 12\% by  
varying the cuts on the three forward detectors
and by using different combinations of two of them. The uncertainty of the 
background which was subtracted statistically in the elastic \cs\ 
was estimated to be 2\% and is included in the total of 12\%. It was 
estimated 
by varying the assumed $1/M_X^k$ dependence of the cross section from
$k$=2 to $k$=2.5 ($k=2.2\pm 0.2$ was measured in \cite{refk}).

\begin{table}[htb]
\begin{center}
\begin{tabular}{|l| c c|}
\hline
              & $\mu^+\mu^-$  & $e^+e^-$ \\
%\hline
\hline
Separation elastic/p.diss. & \multicolumn{2}{c|}{12\%}\\
Trigger               & 9\% & 8\% \\
Single lepton identification & 5\% & 4\% \\
Single track reconstruction & \multicolumn{2}{c|}{3\%}\\
Acceptance                  & \multicolumn{2}{c|}{4\%}\\
Branching ratio        & \multicolumn{2}{c|}{4\%}\\
Photon flux            & \multicolumn{2}{c|}{2\%}\\
 
Luminosity             & \multicolumn{2}{c|}{1.5\%}\\ 
\psiprime\ -background  & \multicolumn{2}{c|}{2\%}\\
\hline
{\bf Total } & 18 \% & 17\% \\
\hline
\end{tabular}
\end{center}
\caption{\sl Systematic errors  }
\label{tserr}
\end{table}

The systematic errors of the track reconstruction efficiency (3\%), 
single muon identification (5\%), electron identification (4\%) and 
trigger efficiency (9\% and 8\% for muon and electron triggers) 
are estimated by comparing the Monte Carlo efficiencies with
the efficiencies determined from the data and using remaining
differences as error.

The systematic error of the angular acceptance 
depends mainly on the energy dependence of the  
$\gamma p$ \cs\ and is estimated by varying 
$\sigma_{\gamma p}\propto W_{\gamma p}^{\delta}$ between 
$0.6\leq  \delta \leq 1.0$ which results in a  4\% error.
The uncertainty in the background due to \psiprime\ cascade decays via \jpsi\ 
accounts for the measurement error on the ratio of \psiprime/$\psi$
production. 
For the evaluation of the \gp\ \cs\ an uncertainty in the photon flux of $2\%$ is estimated by varying the upper limit of the $Q^2$ integration 
by $\pm 1\GeV$.
 
\subsubsection*{{\boldmath $p_t^2$} distribution}

\begin{figure}[tbp]
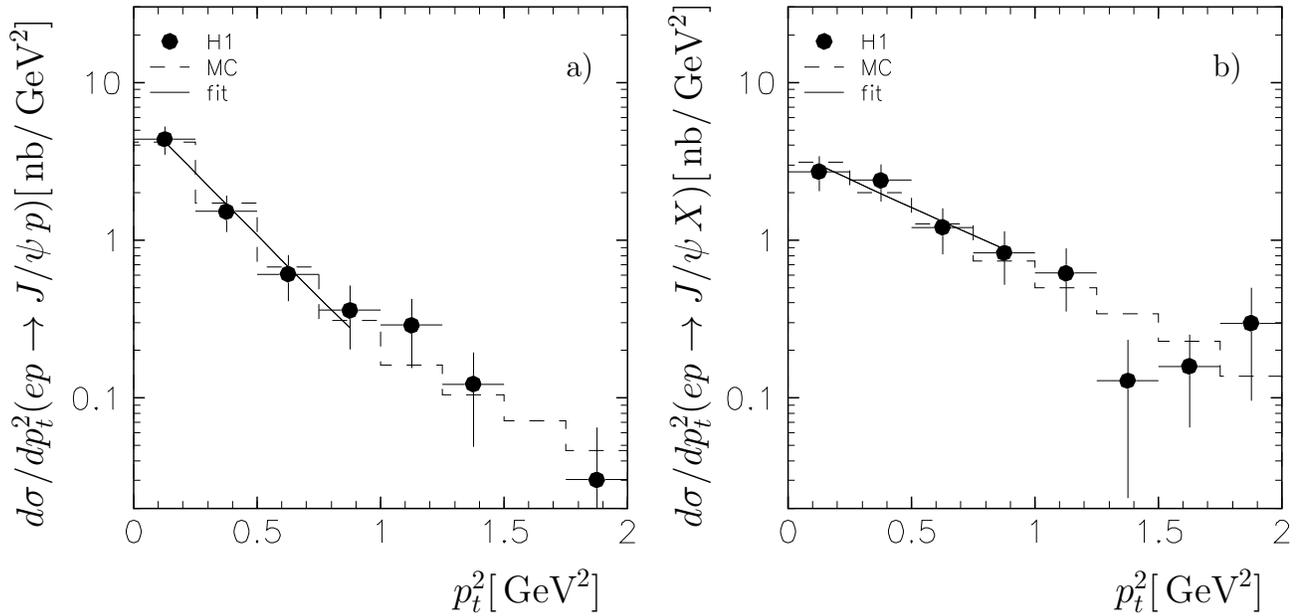

\vspace*{13pt}
\setlength{\unitlength}{1cm}
\begin{picture}(14.0,7.5)
\put(-1.6,-1.6){\epsfig{file=PAPER.ELA.PT2,height=33cm,angle=90}}
\put(0.25,0.5){\begin{rotate}{90}\large $d\sigma/dp_{t}^2
(ep \rightarrow J/\psi \, p)[\nb/\GeV^2]$\end{rotate}}
\put(5.7,-0.5){\large $p_{t}^2[\GeV^2]$}
\put(7.1,-1.6){\epsfig{file=PAPER.PD.PT2,height=33cm,angle=90}}
\put(9.,0.5){\begin{rotate}{90}\large $d\sigma/dp_t^2
(ep \rightarrow J/\psi \, X)[\nb/\GeV^2]$\end{rotate}}
\put(14.5,-0.5){\large $p_t^2 [\GeV^2]$}
\put(7.2,6.5){a)}
\put(15.8,6.5){b)}
\end{picture}
\vspace*{13pt}
\caption{\sl {\bf a)} $d\sigma/dp_{t}^2$ for {\bf elastic} $J/\psi$
production
integrated over $30\GeV\leq W_{\gamma p} \leq 150\GeV$. 
 The error bars
contain statistical and systematic errors added in quadrature.
The straight line is a fit to the data in the range 
$p_t^2\leq 1 \GeV^2$ of the form $\exp(-bp_t^2)$ with $b=4.0\pm0.3~\gmt$. 
The histogram shows the $p_{t}^2$ 
distribution of the DIFFVM Monte Carlo. {\bf b)}
$d\sigma_{\gamma p}/dp_{t}^2$ for $J/\psi$ meson
production with {\bf proton dissociation}
integrated over $30\GeV\leq W_{\gamma p} \leq 150\GeV$. The
straight line is an exponential fit to the data in the range 
$p_t^2\leq 1 \GeV^2$ which yields a value of 
$b=1.6\pm0.3~\gmt$, the dashed line represents the DIFFVM
Monte Carlo calculation. }
\label{ELA.PT2}
\label{PD.PT2}
\end{figure}

The $t$--distribution is one of the distinctive features of a diffractive
process. It is expected to be exponential and the slope
parameter in Regge type models 
is expected to increase with energy (shrinkage) as 
$b=b_0+ 2\alpha'\ln W^2_{\gamma p}/W^2_0$
with $\alpha'=0.25\gmt$ for pomeron exchange~\cite{DL1}.
In contrast, models based on perturbative QCD predict little shrinkage.\\ 
Since the scattered electron is not measured, $t$ can only be approximately
determined for the data assuming $Q^2 = 0$, then $t\approx -p_t^2$
where $p_t^2$ is the transverse momentum of the \jpsi\ with respect to
the beam axis.
The data analysis was carried out as in the previous section, but the 
efficiencies and background proportions were determined in bins of $p_t$. 
The cross section $d\sigma_{ep}/dp_t^2$ is shown in Fig.~\ref{ELA.PT2} a. A 
steep slope is observed for $p_t^2<1\GeV^2$ with a tail 
towards higher values. The tail is compatible with Monte Carlo expectations
for events with $Q^2\neq 0$.

% ELASTIC PT2
%%%%%%%%%%%%%%

The slope obtained by a log-likelihood-fit
of an exponential $\exp{(-b\,p_t^2)}$ 
to the data below $p_t^2\leq 1\GeV^2$ is
\begin{displaymath} 
b=\left(4.0\pm 0.2 \pm 0.2\right)\gmt.
\end{displaymath}

The first error is statistical. The
second one contains the dominant systematic contributions estimated 
by varying the upper limit of the fit region in 
$p_t^2$ between $0.75$ and $1.25 \GeV^2$ or by using a $\chi^2$-fit, 
alternatively. The error in the slope parameter due to using
$p_t^2$ instead of $t$  is calculated in the simulation to be $-10\%$.

For the $W_{\gamma p}$ range from 30 \GeV\ to 90 \GeV\ the fit yields
$b=\left(3.7 \pm 0.3 \pm 0.2\right){\GeV}^{-2}$ and 
$b=\left(4.5 \pm 0.4 \pm 0.3\right){\GeV}^{-2}$ 
for the interval from 90 \GeV\ to 150 \GeV. 
Within the errors no clear
evidence for shrinkage is observed. 
The expected change in the $b$ slope from the Regge prediction is 
only $0.6\GeV^{-2}$ in the HERA energy range. This is of the same order 
of magnitude as the experimental errors, therefore
no conclusion can be drawn with present statistics from the data 
in the HERA energy range alone. 

The experimental situation at low cms energies is unclear: $b$-values  
ranging from $b\approx 3\GeV^{-2} $ to $b\approx 5\GeV^{-2}$ were
measured~\cite{aemc83,cemc92,ee40182,bna14,gnmc94} thus
also preventing any conclusion about shrinkage.

%%%%%%%%%%%%%%%%%%%%%%%%%%%%%%%%%%%%%%%%%%%%%%%%%%%%%%%%%%%%%%%%%%%
%
%   P R O T O N   D I S S O
%
%%%%%%%%%%%%%%%%%%%%%%%%%%%%%%%%%%%%%%%%%%%%%%%%%%%%%%%%%%%%%%%%%%%

\subsection{$J/\psi$ Production with Proton Dissociation}
\label{spdis}
\subsubsection*{Energy dependence and $p_t^2$ distribution}

For the measurement of the proton dissociation cross section the two 
track sample {\em with} a signal in one of 
the forward detectors is used as defined in
section \ref{twotr}. The Monte Carlo generator DIFFVM for proton 
dissociation with parameters as in section \ref{ghghjk} is  used 
for acceptance and efficiency determination.
The procedure for calculating the \cs\ and systematic errors
is the same as in the elastic case. The main contribution to the systematic
error is due to the selection of the event sample with the help of 
the forward detectors. The size of the error is essentially the same 
(12\%) since both \cs s, the elastic and the \pdiss\ \cs, are nearly of 
the same magnitude. The systematic error of the acceptance may be slightly
larger than in the elastic case due uncertainties in the $M_X$ dependence,
this is however not taken into account.

The  \gp\ \cs\ for \jpsi\ production with   
proton dissociation derived from $J/\psi$ decays to
\mm\ and \ee\ as a function of $W_{\gamma p}$  
is given in table~\ref{elacomb} and
shown in Fig. \ref{PD.WGP}b. It can be fitted 
by a $W_{\gamma p}^{\delta}$ dependence with $\delta=1.2\pm0.2$
(statistical and systematic error),
slightly larger than the value for the elastic data,
which may be explained by the increased phase space for producing
excited nucleon systems at higher energy.
%The ratio of the elastic to \pdiss\ \cs\ is......uuuuuuuuuuuuuu

The cross section $d\sigma_{ep}/dp_t^2$ for  $J/\psi$
production with proton dissociation 
is shown in Fig. \ref{PD.PT2} b. A log-likelihood
fit below $1 \GeV^2$ gives: 

\begin{displaymath}
b=\left(1.6 \pm 0.3 \pm 0.1\right){\GeV}^{-2}.
\end{displaymath}

The first error is statistical while the second one represents
the  systematic error estimated as in the elastic case. 
The error in the slope parameter due to using
$p_t^2$ instead of $t$  is calculated in the simulation to be $-6.5\%$. 
The slope parameter is roughly a factor 2 smaller 
than for pure elastic scattering.

\subsubsection*{Decay angular distribution}

The polarisation of the \jpsi\ can be accessed via the angular distribution of
the \jpsi\ decay leptons. 
The angle $\theta^*$ is used, which is the angle
in the \jpsi\ rest frame, between the direction of the positively 
charged decay lepton and the \jpsi\ direction in the \gp\ cms 
(helicity frame)~\cite{Bauer,Schilling_Wolf}.
Assuming s-channel helicity  conservation the expected angular
distribution is:

\begin{displaymath}
\frac{d\sigma}{d\cos\,\theta^*}\propto
(1-\rho)\sin^2\theta^*+\rho\frac{1+\cos^2\theta^*}{2}
\label{costh}
\end{displaymath}

where $\rho$, the fraction of transversely polarised \jpsi\ mesons,
is predicted to be one.

\begin{figure}[tbp]
\vspace*{13pt}
\setlength{\unitlength}{1cm}
\begin{picture}(14.0,9.0)
\put(-.75,-2.0){\epsfig{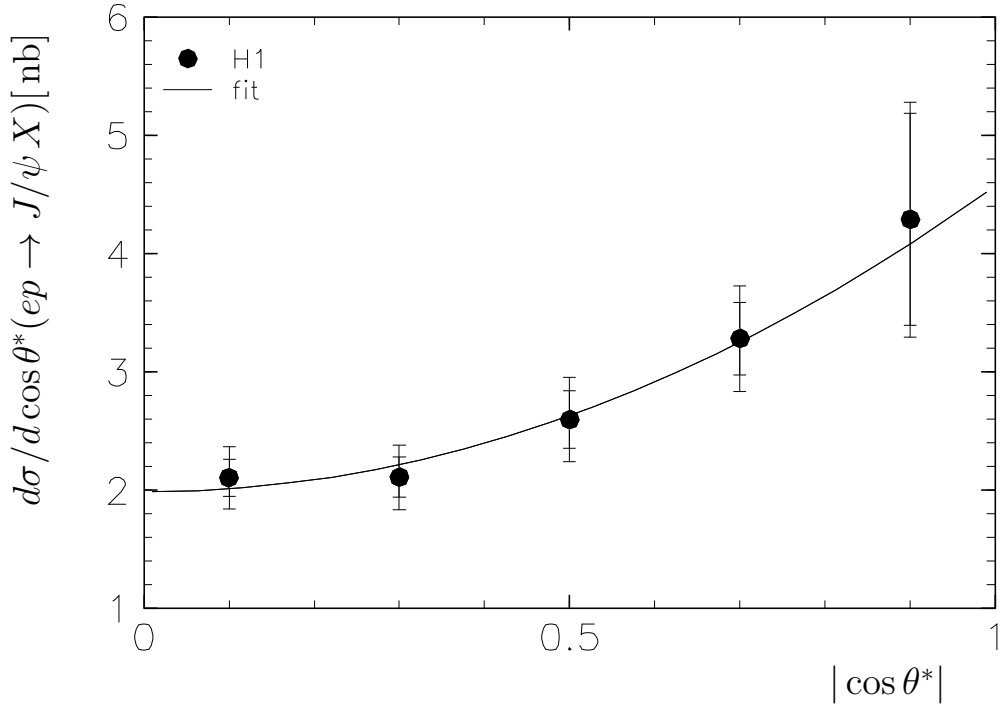}}
\put(1.5,2.0){\begin{rotate}{90}\large $d\sigma/d\cos\theta^*
(ep \rightarrow J/\psi \, X)[\nb]$\end{rotate}}
\put(12.,-0.5){\large $|\cos\theta^*|$}
\end{picture}
\vspace*{13pt}
\caption{\sl $d\sigma/d\cos\theta^*$ for  diffractive $J/\psi$
production (elastic and \protect\pdiss). The inner
error bars  are statistical, the outer ones
contain statistical and systematic errors added in quadrature.
The curve is a fit of the form $\propto (1+\mbox{const}\,\cos^2\theta^*)$.}
\label{COSTHE}
\end{figure}

The $ep$ cross section for both decay channels
\jmm\ and \jee\ as a function of $\cos \theta^*$ is shown in Fig. \ref{COSTHE}.
The full two track sample is used in this analysis including elastic and 
\pdiss\ samples. A $\chi^2$ fit yields $\rho=1.2\pm 0.2$, which is
consistent with s-channel helicity conservation.

 %ELASTIC COSTHE
%%%%%%%%%%%%%%%

%%%%%%%%%%%%%%%%%%%%%%%%%%%%%%%%%%%%%%%%%%%%%%%%%%%%%%%%%%%%%%%%%%%
%
%   I N E L A S T I C
%
%%%%%%%%%%%%%%%%%%%%%%%%%%%%%%%%%%%%%%%%%%%%%%%%%%%%%%%%%%%%%%%%%%%

\section{Inelastic {\boldmath $J/\psi$} Production}
\label{secinela}

\begin{figure}[tbp]
\vspace*{13pt}
\setlength{\unitlength}{1cm}
\begin{picture}(14.0,9.0)
\put(-.75,-2.0){\epsfig{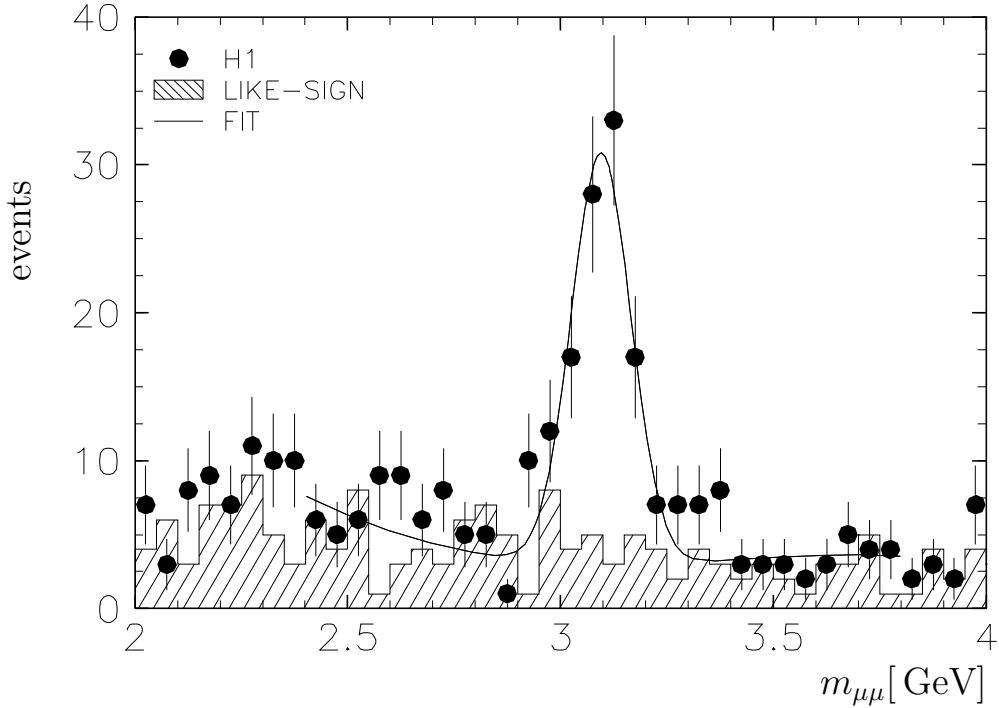}}
\put(1.5,5.){\begin{rotate}{90}\large events\end{rotate}}
\put(12.,-0.5){\large $m_{\mu\mu}[\GeV]$}
\end{picture}
\vspace*{13pt}
\caption{\sl Mass distribution for $\mu^+\mu^-$ of the {\bf inelastic}
         event sample.
         The curve
        is a fit of a Gaussian plus a polynomial background to the
        $J/\psi$ mass region. The shaded histogram shows the 
        mass distribution of like sign muon pairs. The 
        maximum of the fit is at $(3.10 \pm 0.01)\GeV$ with a width
        of 65~MeV. }
\label{INELAMASS}       
\end{figure}

The event selection for inelastic \jpsi\ production starts from the 
preselected lepton pair sample described in section~\ref{presel}. 
The mass distribution for \mm\ pairs with at least one additional track
from the interaction point is shown in Fig.~\ref{INELAMASS}. 
The inelastic \jpsi\ candidates are selected by a cut around the nominal
mass 
of $\pm 225 \MeV$. The background is due to muons from leptonic 
$\pi$ and $K$ decays 
and misidentified hadrons. It is subtracted by using the data above and below
the mass peak as an estimate of the background. 
 
%MASS INEL
%%%%%%%%%%

% Z
%%%%%%

\begin{figure}[tbp]
\vspace*{13pt}
\setlength{\unitlength}{1cm}
\begin{picture}(14.0,9.0)
\put(-0.75,-2.0){\epsfig{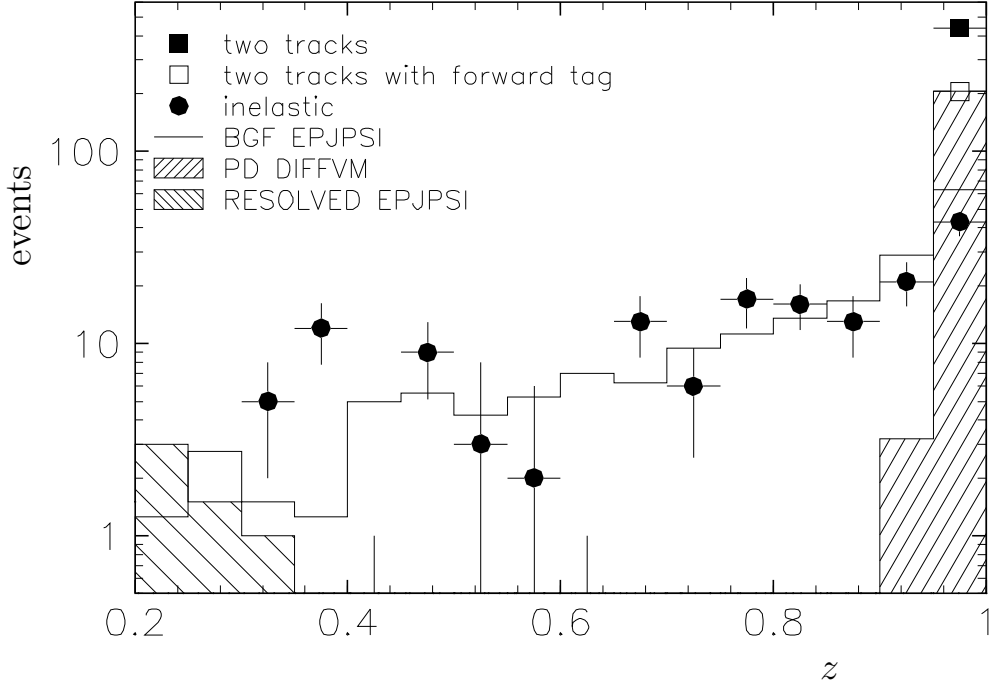}}
\put(1.5,5.){\begin{rotate}{90}\large events\end{rotate}}
\put(12.,-0.5){{\large $z$}}
\end{picture}
\vspace*{13pt}
\caption{\sl Distribution of the elasticity $z$ for  $J/\psi\rightarrow
\mu^+\mu^-$.  The H1 data in the $J/\psi$ mass window are corrected for 
background below the mass peak.
They  are classified as events with exactly two  tracks 
irrespective of any forward tag (full square), events with 
exactly two tracks and a forward tag (open square) and inelastic
events  with $z>0.3$ (full circle).
For comparison the following model calculations are shown: 
a diffractive model including 
proton dissociation (DIFFVM) normalized to the 
data (two tracks with forward tag);
a LO photon gluon fusion
model (LO BGF EPJPSI) for inelastic $J/\psi$ production using the gluon 
distribution from the MRSD-' parametrisation, and
the resolved contribution calculated in EPJPSI with the GRV-LO photon
structure function.
The resolved contribution is scaled with the same normalisation
factor as the contribution from  photon gluon fusion.}
\label{ZSTUDIES}        
\end{figure}

The elasticity $z$ -- calculated according to eqn. \ref{yjbz} in
section~\ref{formulas} -- is used to define the inelastic sample. 
In Fig. \ref{ZSTUDIES} an overview over {\em all} candidates for \jmm\ 
is shown as function of $z$.
The measured $z$ distribution is shown for the 
inelastic \jpsi\ candidates with $z>0.3$. 
The total two track sample and also the subsample with activity in the 
forward detectors which were used in the previous section to evaluate
the elastic and \pdiss\ \cs s, respectively, are shown in the highest 
$z$-bin of the figure.
The data in  Fig. \ref{ZSTUDIES} are compared with three Monte Carlo models:
the contribution of photon gluon fusion as 
modelled in the generator EPJPSI (LO \colsing\ model)
is shown, at high $z$ the 
contribution of \pdiss\ processes as modelled in DIFFVM can be seen, and 
at low $z$ the contribution of the hadronic component of the photon, 
the ``resolved photon process~\cite{resolved}'', is shown as modelled by 
EPJPSI. Note that the 
simulations are normalized to the data. 
DIFFVM with \pdiss\ is normalised to the 
two track data with forward detector signal. The contribution of 
photon gluon fusion is normalised to the inelastic data for $0.45<z<0.90$
and the resolved component is multiplied by the same normalisation factor.

An inelastic sample is
selected by requiring $0.45 \leq z \leq 0.90$ which 
rejects diffractive and resolved photon events and yields a total of
$\approx 100$ \jpsi\ candidates.
The  data sample is corrected for efficiency and 
acceptance using the Monte Carlo generator EPJPSI~\cite{jungws} with the 
parametrisation of the gluon density of MRSD-'~\cite{mrsd-}.
The \cs\ for inelastic \jpsi\ production is determined for $0\leq z\leq0.90$
in order to compare with theory. Therefore 
a correction is applied for the loss below $z<0.45$, which is 
estimated to $13.5\pm 2\%$, where the 
error is due to using a different gluon distribution (MRSD0'). 
An additional correction takes into account the contamination by 
$J/\psi$ meson production via the resolved photon process.
This background is 
estimated  using an option of the Monte Carlo generator EPJPSI.
Using the 
GRV-LO-parametrisation~\cite{pGRV} of the photon structure function yields a 
background estimate of 2.5\% for $z>0.45$. 
The contribution changes by $+ 2\%$ 
using as photon structure function the parametrisation of 
LAC1~\cite{pLAC1}.

Systematic errors for trigger and selection efficiency  
are estimated as in the elastic analysis.
In addition to the relevant errors quoted in table~\ref{tserr} which yield a 
contribution of 12\%, 
a contribution to the systematic error of 10\% is estimated 
specifically for the inelastic \cs.  
This number contains the uncertainty
due to background subtraction, an error of the contamination
with resolved events, the correction for the
low $z$ region. A further contribution to the error of 5\% is due to 
a systematic shift in the $z$ determination after reconstruction.
In the simulation $z$ is  found to be
systematically  shifted  to higher values after reconstruction 
with respect to the generated value
by 4\% for $z=0.8$ and by 10\% for $z=0.5$.

\subsubsection*{Energy dependence of $\gamma p$ cross section}
\label{secinelw}

\begin{figure}[tbp]
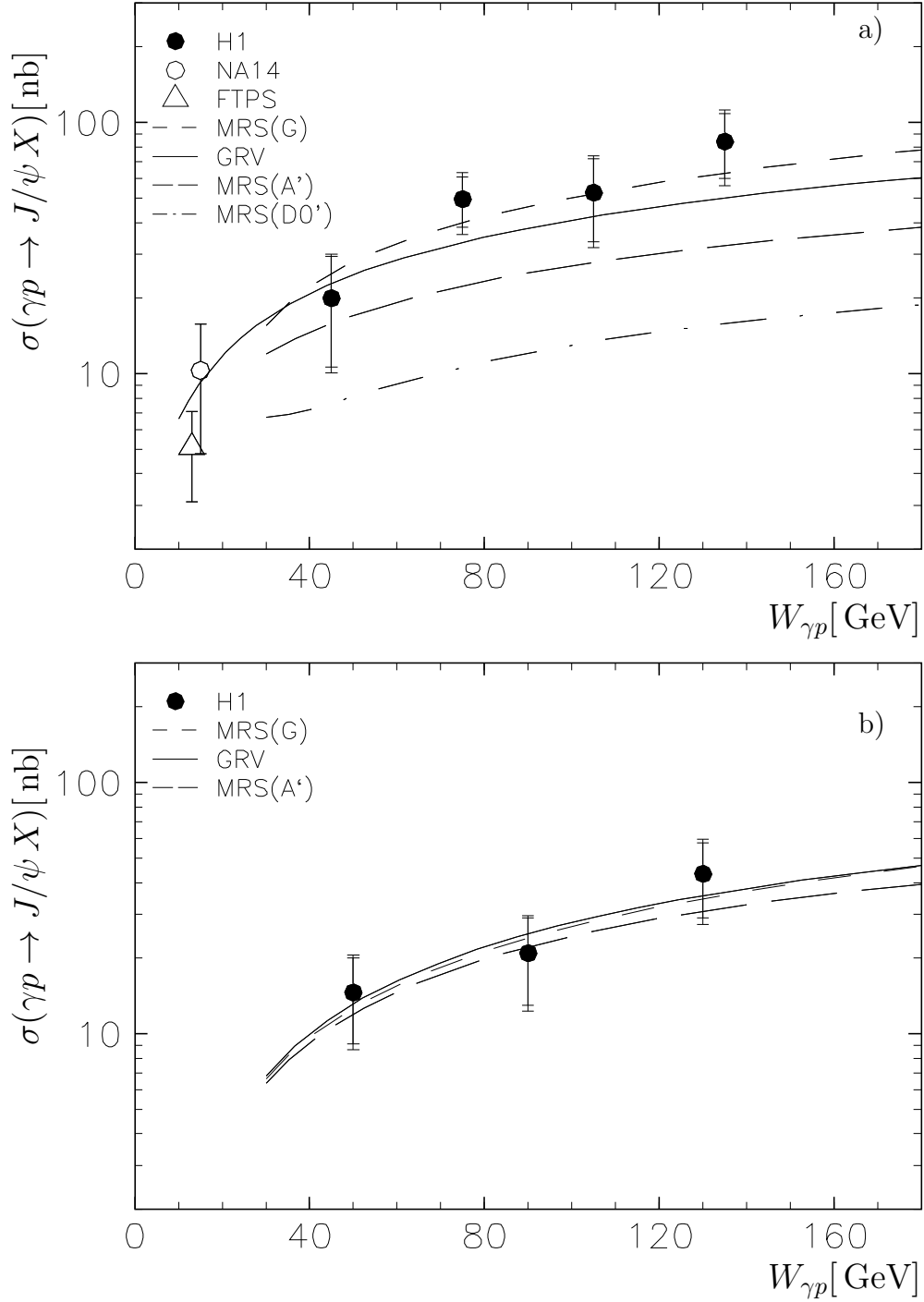

\vspace*{13pt}
\setlength{\unitlength}{1cm}
\begin{picture}(14.0,16.0)
\put(-0.75,7.5){\epsfig{file=PAPER.INEL.WGP1,height=35cm,angle=90}}
\put(1.5,13.0){\begin{rotate}{90}\large $\sigma(\gamma p \rightarrow J/\psi \, X)[\nb]$\end{rotate}}
\put(12.,9.){\large $W_{\gamma p}[\GeV]$}
\put(-0.75,-2.0){\epsfig{file=PAPER.INEL.WGP2,height=35cm,angle=90}}
\put(1.5,3.0){\begin{rotate}{90}\large $\sigma(\gamma p \rightarrow J/\psi \, X)[\nb]$\end{rotate}}
\put(12.,-0.5){\large $W_{\gamma p}[\GeV]$}
\put(13.3,17.5){a)}
\put(13.3,7.5){b)}
\end{picture}
\vspace*{13pt}
\caption{\sl Total cross section for {\bf inelastic} $J/\psi$-production for
$z<0.9$ (a) and $z<0.8$ and $p_t^2>1\GeV^2$ (b). 
The inner error bars of the H1 points are statistical, the outer
include statistical and
systematic errors added in quadrature. The
curves represent NLO calculations~\protect\cite{kraem2} for different gluon distributions
and contain a 15\% correction taking into account \psiprime\ background.} 
\label{INELA.WGP1}
\end{figure}

The $\gamma p$ cross section for inelastic $J/\psi$ production
can be derived in a similar way as in 
section \ref{sect elas}. Efficiencies, event numbers
and the final cross section for $z<0.9$ are given in table~\ref{inelamu}.
The resulting $\gamma p$ cross section is shown in Fig. \ref{INELA.WGP1}a.
The data are compared to calculations in the colour singlet model 
including NLO contributions~\cite{kraem2}.

%%%% I N E L A S T I C

\begin{table}[tb]
\begin{center}
\begin{tabular}{|l|c|c|c|c|}
\hline 

\multicolumn{5}{|c|}{$J/\psi \rightarrow \mu^+\mu^-$}\\
\hline 
 $ W_{\gamma p}[\GeV]$& 30--60 &60--90 & 90--120 & 120--150\\ 
\hline 
 events & & & & \\ 
(bg. corrected)& \rb{8.0$\pm$3.7} &\rb{ 27.0$\pm$6.1} &\rb{ 23.0$\pm$8.3} 
                 &\rb{ 27.0$\pm$7.8}\\
\hline 
$\epsilon_{acc}$&0.326$\pm$0.013  & 0.799$\pm$0.032 &
0.854$\pm$0.034 & 0.731$\pm$0.029  \\ 
$\epsilon_{analysis}$&0.333$\pm$0.027 & 0.387$\pm$0.031 & 0.329
$\pm$0.026 
& 0.279$\pm$0.022 \\ 
$\epsilon_{trigger}$&0.304$\pm$0.024  & 0.290$\pm$0.023 & 0.413
$\pm$0.033 
& 0.646$\pm$0.052 \\ 
\hline 
\seppsix [nb] &1.5$\pm$0.7$\pm$0.2 & 1.8$\pm$0.4$\pm$0.3 
& 1.2$\pm$0.4$\pm$0.2 & 1.3$\pm$0.4$\pm$0.2 \\ 
\hline 
$\Phi_{\gamma /e}$ &0.0736 & 0.0370 & 0.0229 & 0.0149 \\ 
\hline 
\sgppsix [nb] &20.0$\pm$9.4$\pm$3.2 & 49.5$\pm$11.1
$\pm$7.9 
& 52.6$\pm$19.0$\pm$8.4 & 84.1$\pm$24.3$\pm$13.4 \\ 
\hline 
\end{tabular}
\end{center}
\caption{\sl {\bf Inelastic {\boldmath  $J/\psi$} production}: 
Efficiencies and cross sections for inelastic 
$J/\psi \rightarrow \mu^+ \mu^-$ production with $0<z<0.90$.
The number of events is background corrected.
$\Phi_{\gamma/e}$ is the photonflux integrated over $Q^2$ and \protect\wgp.
The errors of acceptance, analysis and trigger efficiency are
statistical and systematic errors added in quadrature.
The first error of the cross
sections is statistical, the second is systematic.}
\label{inelamu}
\end{table}

%  WGP INELASTIC 
%%%%%%%%%%%%%%%%%%%%%

As input the theoretical calculations use the charm mass $m_c=1.4\, \GeV$,
 $\Lambda_{\overline{MS}}=300 \MeV$, the renormalisation scale is 
chosen to be identical
 to the factorisation scale, $\sqrt{2}\, m_c$. Predictions  are
 shown for different gluon density 
distributions that have been derived from recent data.
The gluon distributions differ in the low $x$ behaviour
where $x$ is the fraction of the proton momentum carried by the gluon. The 
distributions used can at low $x$ be parametrised as 
$x\,g(x)\propto x^{-\lambda}$, and the values for $\lambda$ range between 0
(MRSD0')~\cite{mrsd-}) and 0.4 (MRSG~\cite{MRS}).
In Fig.~\ref{INELA.WGP1}a the curves contain an additional contribution
of inelastic \psiprime\ production ($\approx 15\%$, 
estimated in~\cite{kraem2}).
They reproduce approximately the energy behaviour of the data,
but cover a wide range in absolute normalisation. The agreement with the data 
becomes better with increasing steepness of the gluon density at low $x$.

\begin{figure}[tbp]
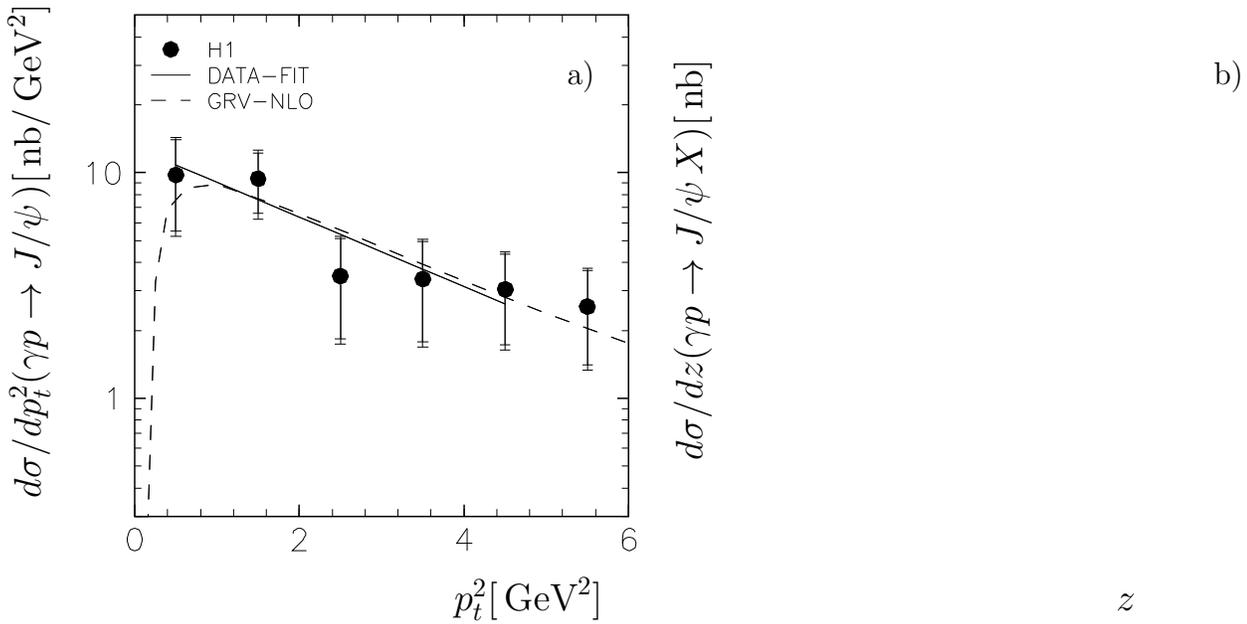

\vspace*{13pt}
\setlength{\unitlength}{1cm}
\begin{picture}(14.0,7.5)
\put(-1.6,-1.6){\epsfig{file=PAPER.INEL.PT2,height=33cm,angle=90}}
\put(.2,1.0){\begin{rotate}{90}\large $d\sigma/dp_t^2(\gamma p \rightarrow
J/\psi \, )[\nb/\GeV^2]$\end{rotate}}
\put(5.7,-0.5){\large $p_t^2 [\GeV^2]$}
\put(7.2,6.5){a)}
\put(15.8,6.5){b)}
\put(7.1,-1.6){\epsfig{file=PAPER.INEL.Z,height=33cm,angle=90}}
\put(9.0,1.5){\begin{rotate}{90}\large $d\sigma/dz(\gamma p \rightarrow J/\psi
\, X)[\nb]$\end{rotate}}
\put(14.5,-0.5){{\large $z$}}
\end{picture}
\vspace*{13pt}
\caption{\sl a) $d\sigma_{\gamma p}/dp_{t}^2$ for {\bf inelastic} $J/\psi$
production with $z<0.9$ 
integrated over $30\GeV\leq W_{\gamma p} \leq 150\GeV$. The
straight line is an exponential fit to the data in the range 
$p_t^2\leq 5 \GeV^2$, the dashed line is a NLO calculation~\protect\cite{kraem2}
at $W_{\gamma p}=100 \GeV$
with the GRV structure function. 
b) $d\sigma_{\gamma p}/dz$ for inelastic $J/\psi$ production
with a cut $p_t^2\geq 1\GeV^2$, 
in comparison with NLO calculations in the colour singlet model   
with the GRV structure function. The LO colour octet calculation is
from ~\protect\cite{cacci}.}
\label{INELA.PT2}
\label{INELA.Z}
\end{figure}

The NLO calculation is not fully under control for 
$p_t^2\ra 0$  and $z\ra 1$~\cite{kraem3}. 
Missing contributions of even higher order
cause problems as can be seen in Fig.~\ref{INELA.PT2}
where the theoretical  $p_t^2$ distribution  bends over 
for $p_t^2\lsim 1\GeV^2$. 

%
% PT2 INELASTIC
%%%%%%%%%%%%%%%%

Therefore the data and prediction are shown  in Fig.~\ref{INELA.WGP1}b 
for a restricted kinematical range, $z<0.8$ and
$p_t^2>1\GeV^2$ and thus requiring the emitted gluon to be hard 
(thereby reducing the data sample by a factor 2).
Improved agreement is observed with all gluon density distributions 
but
the sensitivity is reduced.

\subsubsection*{\boldmath{$p_t^2$, $z$-distribution}}
 
In Fig.~\ref{INELA.PT2}a the distribution of the transverse momentum $p_t^2$ 
of the \jpsi\ is shown 
for $z < 0.9$. A log-likelihood fit to the data  of an exponential 
$\exp{(-b\,p_t^2)}$ in the region $p_t^2\leq 5\GeV^2$
yields:

\begin{displaymath} 
b=\left(0.39 \pm 0.06 \pm 0.03 \right){\GeV}^{-2}
\end{displaymath}

where the first error is statistical and where the systematical error 
is estimated by
using a a $\chi^2$-fit and varying the fit region.
This result is in agreement with the NLO-calculation~\cite{kraem2} which
predicts  a slope
of $b= 0.3 {\GeV}^{-2}$ above a $p_t^2>1\GeVt$.
       
The  differential energy distribution of the \jpsi, $d\sigma/dz$, 
is shown in Fig.\ref{INELA.Z}b with a cut in $p_t^2>1\GeVt$.
It is
compared with the NLO calculation~\cite{kraem2} in the colour singlet model
with the GRV~\cite{GRV} structure function for the proton. 
Agreement in shape and normalisation is found within errors.
This is interesting in view of the speculations about possible additional
colour octet contributions to photoproduction of \jpsi. These yield a 
large inelastic contribution at $z>0.8$~\cite{cacci} after fixing the 
normalisation by the CDF data\cite{teva}. The strong increase 
indicated in Fig.~\ref{INELA.Z}b is not supported by the present data.

%%%%%%%%%%%%%%%%%%%%%%%%%%%%%%%%%%%%%%%%%%%%%%%%%%%%%%%%%%%%%%%%%%%
%
%   PSI PRIME
%
%%%%%%%%%%%%%%%%%%%%%%%%%%%%%%%%%%%%%%%%%%%%%%%%%%%%%%%%%%%%%%%%%%%

\section{{\boldmath $\psi ^{\prime}$} Production}

\label{psiprime}
 
The same data sample as for the $J/\psi$ analysis
is used to measure the cross section 
for the diffractive photoproduction of $\psi ^{\prime}$ mesons
($\psi(2S)$), by 
searching for the decay $\psi^{\prime} \rightarrow J/\psi 
\pi ^{+} \pi ^{-}$, where the $J/\psi$ subsequently decays to
a muon pair. 

For this investigation, only the muon triggers are utilised, but
the identification of the muon pair is otherwise identical to
that used in the $J/\psi$ analysis. It is further required that,
in addition to the muon pair, exactly two other, oppositely
charged tracks are found in the central tracking detector which are assumed to
be pions. For events in which both pion candidates are measured in the
central tracking detector to have a transverse momentum greater
than 150 \MeV, a mass difference is formed, defined as 
$\Delta M = m_{\mu ^{+} \mu ^{-} \pi ^{+} \pi ^{-}} - m_{\mu ^{+} \mu ^{-}}$. 
A peak of 7 events is found in an interval of $\pm 60\,\MeV$
around the mass difference,
$\Delta M = m_{\psi ^{\prime}} - m_{J/\psi}$, with no 
background events seen outside the peak.
 
Monte Carlo studies show that this
selection restricts the kinematic domain of acceptance to
$z' > 0.95$, $40 \GeV < W_{\gamma p} < 160 \GeV$ and 
$Q ^{2} < 4 \GeV^{2}$, where $z'=E_{\psi'}/E_{\gamma}$ in the proton rest frame.
The acceptance is estimated using the 
same Monte Carlo generator as was used for the $J/\psi$ analysis,
with parameters of the model chosen to match the $J/\psi$ data.
A  mixture of elastic and proton dissociation events is
assumed, which is compatible with that found in the $J/\psi$
events. No separation of elastic events
from events with proton dissociation is attempted.

In the kinematic region above, the combined acceptance of trigger
and selection is found to be $4.4 \pm 0.9\%$. Assuming a branching
ratio for this $\psi ^{\prime}$ decay of $0.019 \pm 0.002$~\cite{pdg}, the
$ep$ cross section for the production of $\psi ^{\prime}$ in the
kinematic domain $z' > 0.95$, $40 \GeV < W_{\gamma p} < 160 \GeV$ and 
$Q ^{2} < 4 \GeV^{2}$ is $2.9 \pm 1.1 \pm 0.6 \nb$, where the first
error given is statistical, the second systematic. This translates
into a photoproduction cross section,
$\sigma _{\gamma p}(W_{\gamma p}=80\GeV) = 24 \pm 9 \pm 5\nb$ for
$\psi ^{\prime}$ production with $z' > 0.95$ and $Q ^{2} < 4 \GeV^{2}$.
The corresponding cross section for $J/\psi$ production
is the diffractive \cs\  (including elastic and proton dissociation)
which at \wgp$=80\GeV$ 
is $\sigma _{\gamma p} = 119.5 \pm 11.2 \pm 12.0 \nb$.
The ratio of $\psi'$ to $J/\psi$ production is then $0.20\pm 0.09$
which is in agreement with previous 
measurements~\cite{aemc83,bna14,he40183}.

%%%%%%%%%%%%%%%%%%%%%%%%%%%%%%%%%%%%%%%%%%%%%%%%%%%%%%%%%%%%%%%%%%%
%
%   c o n c l u s i o n
%
%%%%%%%%%%%%%%%%%%%%%%%%%%%%%%%%%%%%%%%%%%%%%%%%%%%%%%%%%%%%%%%%%%%

\section{Summary}

\jpsi\ meson production in the photoproduction limit is analysed in the
elastic channel. Corrections for contributions from events with proton
dissociation are applied almost entirely
on an event to event basis.
The following results are found:

\begin{itemize}
\item
The total $\gamma p$ cross section is observed to increase with energy  as 
$W_{\gamma p}^{\delta}$ with $\delta=0.64 \pm 0.13$ for H1 data alone, 
and $\delta=0.90 \pm 0.06$
including the ZEUS and low energy data.
 This increase is faster than predicted in soft 
diffractive models ($\delta=0.32$).
\item A calculation using the model by Ryskin in the framework of perturbative 
QCD results in a good description 
of the energy dependence if the gluon distribution from the set MRSA' 
is used.

\item The $p_t^2$ distribution of the elastic \jpsi\ data below $1\GeV^2$ 
can be fitted with an exponential $e^{-bp_t^2}$, 
$b=4.0\pm 0.2\pm 0.2\GeV^{-2}$. 
\end{itemize}

The $\gamma p$ cross cross section for $J/\psi$ meson production with proton 
dissociation was determined. The \gp\ \cs\ is as large as the elastic one and 
its energy dependence is 
slightly steeper than the elastic \cs, $\delta=1.2\pm0.2$.
The $p_t^2$ distribution
is flatter than for the elastic process. The slope parameter is
approximately a factor 2 smaller than for elastic data. 
 
The 
angular distribution in the helicity frame was determined for 
the complete diffractive data sample, i.e. including elastic and \pdiss\ 
contributions. The distribution in the helicity frame 
is consistent with  s-channel
helicity conservation, $\rho=1.2\pm 0.2$, where $\rho$ is
the fraction of transversely polarised \jpsi\ mesons.

Inelastic $J/\psi$ meson production can be described well by
a QCD calculation in NLO of the 
colour singlet model 
if cuts are used which demand the emitted gluon to be hard,
 $p_t>1\GeV$ and $z<0.8$, in order to make perturbation theory  applicable. 
The agreement is observed in the shape
of the $W_{\gamma p}$, $p_t^2$ and $z$ distributions. The absolute
normalization between data and theory agrees within the experimental
errors of $\approx 30\%$, 
using any gluon distribution function which describes the
$F_2$ data at low $x$. %\cite{h1f2}. 
A comparison to the relative energy distribution $z$ of inelastic \jpsi\ 
production with $p_t^2>1\GeVt$ calculated 
in a \coloct\ model shows disagreement at high values and
makes a large \coloct\ contribution unlikely.

Diffractive production of \psiprime\ was observed. An estimate of the ratio
of the \cs\ to the one for \jpsi\ yields $0.20\pm 0.09$ in agreement with
previous measurements at lower energy.
  
\subsection*{Acknowledgements}
We are very grateful to the HERA machine group whose outstanding
efforts made this experiment possible. We acknowledge the support
of the DESY technical staff. We appreciate the big effort of the
engineers and technicians who constructed and maintained the
detector. We thank the funding agencies for financial support
of this experiment. We wish to thank the DESY directorate for
the support and hospitality extended to the non--DESY members
of the collaboration.
We wish to thank M. Kr\"amer from the DESY theory group
for many enlightening discussions and good collaboration. Thanks to R.
Roberts (RAL) for providing  us with his numerical results for elastic 
\jpsi\ production.

 \end{document}